\documentclass[fleqn,usenatbib]{mnras}

\usepackage{newtxtext,newtxmath}
\usepackage[T1]{fontenc}

\DeclareRobustCommand{\VAN}[3]{#2}
\let\VANthebibliography\thebibliography
\def\thebibliography{\DeclareRobustCommand{\VAN}[3]{##3}\VANthebibliography}

\usepackage{graphicx}	% Including figure files
\usepackage{amsmath}	% Advanced maths commands
\usepackage{xcolor}     % For comments with different colours
\newcommand{\arsco}{AR~Sco}

\newcommand{\hzs}{Hz~s$^{-1}$}
\title[Photometry and spectroscopy of AR~Sco]{Long-term photometric monitoring and spectroscopy of the white dwarf pulsar AR~Scorpii}

\author[Pelisoli et al.]{
Ingrid Pelisoli,$^{1}$\thanks{E-mail: ingrid.pelisoli@warwick.ac.uk}
T.~R.\ Marsh,$^{1}$
S.~G. Parsons,$^{2}$
A.\ Aungwerojwit$^{3}$,
R.~P.\ Ashley,$^{4}$
E. Breedt,$^{5}$\newauthor
A.~J. Brown,$^{2}$
V.~S.\ Dhillon,$^{2,6}$
M.~J. Dyer,$^{2}$
M.~J. Green,$^{7}$
P. Kerry,$^{2}$
S.~P. Littlefair,$^{2}$\newauthor
D.~I. Sahman,$^{2}$
T. Shahbaz,$^{6,8}$
J.~F. Wild,$^{2}$
A. Chakpor,$^{9}$
R. Lakhom$^{9}$
\\
$^{1}$Department of Physics, University of Warwick, Gibbet Hill Road, Coventry, CV4 7AL, UK\\
$^{2}$Department of Physics and Astronomy, Hicks Building, The University of Sheffield, Sheffield, S3 7RH, UK\\
$^{3}$Department of Physics, Faculty of Science, Naresuan University, Phitsanulok, 65000, Thailand\\
$^{4}$Isaac Newton Group of Telescopes, Apartado de Correos 321, Santa Cruz de La Palma, E-38700, Spain\\
$^{5}$Institute of Astronomy, University of Cambridge, Madingley Road, Cambridge CB3 0HA, UK\\
$^{6}$Instituto de Astrof\'\i{}sica de Canarias (IAC), E-38205 La Laguna,  Tenerife, Spain \\
$^{7}$Department of Astrophysics, School of Physics and Astronomy, Tel Aviv University, Tel Aviv 6997801, Israel\\
$^{8}$Departamento de Astrof\'{i}sica, Universidad de La Laguna (ULL), E-38206 La Laguna, Tenerife, Spain\\
$^{9}$National Astronomical Research Institute of Thailand, 260 Moo 4, T. Donkaew, A. Maerim, Chiangmai, 50180, Thailand\\
}

\date{Accepted XXX. Received YYY; in original form ZZZ}

\pubyear{2022}

\begin{document}
\label{firstpage}
\pagerange{\pageref{firstpage}--\pageref{lastpage}}
\maketitle

\begin{abstract}
AR Scorpii (\arsco) is the only radio-pulsing white dwarf known to date. It shows a broad-band spectrum extending from radio to X-rays whose luminosity cannot be explained by thermal emission from the system components alone, and is instead explained through synchrotron emission powered by the spin-down of the white dwarf.
We analysed NTT/ULTRACAM, TNT/ULTRASPEC, and GTC/HiPERCAM high-speed photometric data for AR~Sco spanning almost seven years and obtained a precise estimate of the spin frequency derivative, now confirmed with 50-$\sigma$ significance. Using archival photometry, we show that the spin down rate of $P/\dot{P} = 5.6 \times 10^6$~years has remained constant since 2005. As well as employing the method of pulse-arrival time fitting used for previous estimates, we also found a consistent value via traditional Fourier analysis for the first time.
In addition, we obtained optical time-resolved spectra with WHT/ISIS and VLT/X-shooter. We performed modulated Doppler tomography for the first time for the system, finding evidence of emission modulated on the orbital period. We have also estimated the projected rotational velocity of the M-dwarf as a function of orbital period and found that it must be close to Roche lobe filling.
Our findings provide further constraints for modelling this unique system.
\end{abstract}


%%%%%%%%%%%%%%%%%%%%%%%%%%%%%%%%%%%%%%%%%%%%%%%%%%

%%%%%%%%%%%%%%%%% BODY OF PAPER %%%%%%%%%%%%%%%%%%

\section{Introduction}

\begin{figure}
	\includegraphics[width=\columnwidth]{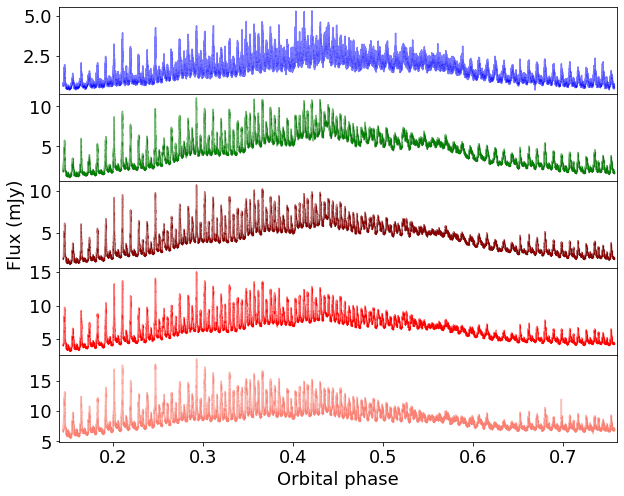}
    \caption{Light curve obtained with HiPERCAM on 2018 April 15 (see Sec.~\ref{sec:phot}). HiPERCAM filters $u, g, r, i, z$ are shown from top to bottom; the y-scale is the same in all panels. Strong pulses can be seen on all bands, on top of the orbital modulation.
    }
    \label{fig:hcam}
\end{figure}

% General introductory paragraph
The binary system AR~Scorpii (henceforth \arsco) is composed of a white dwarf with an M-dwarf companion in a 3.56-hour orbit \citep{Marsh2016}. In contrast to cataclysmic variable systems (CVs), which show this same binary configuration, there is no evidence that the white dwarf in \arsco\ accretes mass from its companion. It is further distinguished from other similar binaries by strongly pulsed emission with a period of 1.97~minutes (see Fig.~\ref{fig:hcam}) over a broad range of wavelengths, from radio \citep{Stanway2018} to X-rays \citep{Takata2018}, which led to it being dubbed a white dwarf pulsar \citep{Geng2016, Buckley2017}. The physical process behind \arsco's pulses is, nonetheless, not the same as for traditional neutron-star radio pulsars \citep{Katz2017, Lyutikov2020}. This is inferred from the fact that no phase coherent radio emission has been reported for \arsco, and from the companion being within the radius of the white dwarf's light cylinder.

% Photometric characterisation
The pulsed variability of \arsco\ was first reported by \citet{Marsh2016}, who detected two components of very similar frequency in \arsco's amplitude spectra, at $\approx 731$ and $\approx 738$ cycles per day (c/d). The difference between the two values corresponds to the orbital frequency, suggesting that the higher frequency component represents the spin of the white dwarf, whereas the lower frequency is a re-processed beat frequency. \citet{Marsh2016} also found the maximum luminosity of \arsco\ to be well in excess of the combined luminosity of the stellar components. Given the lack of accretion signatures (such as strong X-ray luminosity, broad emission lines, or flickering), they proposed this excess luminosity to be powered by the spin-down of the rapidly-rotating white dwarf. They estimated a spin frequency derivative of $\dot{\nu}_\mathrm{spin} = (-2.86\pm0.36) \times 10^{-17}$~\hzs, corresponding to a rate of loss of rotational energy of $L_{\dot{\nu}} = 1.5 \times 10^{26}$~W.

\citet{PotterBuckley2018Fourier} carried out a Fourier analysis of data spanning three years (2015 to 2017) but could not confirm the spin-down found by \citet{Marsh2016}. Instead they found a spin frequency derivative consistent with zero, but were only able to constrain it to the broad range $-20 \times 10^{-17}\,$\hzs to $+10\times 10^{-17}$~\hzs. On the other hand, \citet{Stiller2018} confirmed the occurrence of spin-down, again from just three years  coverage (2016 to 2018), but found that it had a significantly larger magnitude, $\dot{\nu}_\mathrm{spin} = (-5.14\pm0.32) \times 10^{-17}$~\hzs, than determined by \citet{Marsh2016}. \citet{Stiller2018} argued that the apparent inconsistency between the results of \citet{Marsh2016} and \citet{PotterBuckley2018Fourier} arose from an underestimate of $\dot{\nu}_\mathrm{spin}$ in \citet{Marsh2016}, likely due to sparse sampling and low time resolution of their data. They also pointed out that their value of $\dot{\nu}_\mathrm{spin}$ could explain the difference in the spin periods found by \citet{Marsh2016} and \citet{PotterBuckley2018Fourier}, being consistent with the spin period found in both works. More recently, the spin-down rate was further refined to $\dot{\nu}_\mathrm{spin} = (-4.82\pm0.18) \times 10^{-17}$~\hzs by \citet{Gaibor2020}. The spin-down rate measurements have thus converged, with the most recent consistent with each other, but questions remain over the earlier measurements. \citet{Marsh2016} based their measurement upon data from the CRTS survey that preceded (by several years) the more recent intensive follow-up data used by \citet{PotterBuckley2018Fourier}, \citet{Stiller2018} and \citet{Gaibor2020}. The CRTS data were very sparse, and used relatively long exposures (30~secs), which rightly raises a question mark over the reliability of \citet{Marsh2016}'s measurement of spin-down \citep{PotterBuckley2018Fourier}. However, could \citet{Marsh2016}'s result, together with those of \cite{Stiller2018} and \cite{Gaibor2020}, point towards a genuine change in the rate of spin down in AR~Sco? Such behaviour is often seen amongst close relatives of AR~Sco, the intermediate polars \citep[IPs, see e.g. fig. 2 of][]{Patterson2020}. If the same applied to AR~Sco, it could suggest that (fluctuating) mass accretion has some role to play in the spin-down, despite a relative lack of any other evidence for accretion in the system \citep{Marsh2016}. A second puzzle is why \cite{PotterBuckley2018} obtained so loose a constraint upon the spin derivative, some 30 times worse than \cite{Stiller2018} from data covering the same time span. Is the Fourier-based method somehow flawed in the case of AR~Sco?

The orbital period of the system is also expected to change, as a result of angular momentum losses caused by the interplay between magnetic braking and gravitational radiation. \citet{Stiller2018} constrained the orbital frequency derivative to $\dot{\nu}_\mathrm{orbit} \lesssim 2 \times 10^{-18}$~\hzs, which at their precision implied $\dot{\nu}_\mathrm{spin} = \dot{\nu}_\mathrm{beat}$. An even tighter constraint of $\dot{\nu}_\mathrm{orbit} \lesssim 3.8 \times 10^{-20}$~\hzs was found by \citet{Peterson2019} using archival photometric plate data dating back to 1902. They pointed out that their constraint is consistent with the semi-empirical model of \citet{Knigge2011} describing angular momentum loss in CVs, which predicts an orbital frequency derivative of $\dot{\nu}_\mathrm{orbit} \lesssim 2 \times 10^{-20}$~\hzs for \arsco's orbital period.
%Does Knigge's model even apply to AR Sco? Likely no as it doesn't include the WD's magnetic field
Additional studies of the orbital behaviour were performed by \citet{Littlefield2017}, who found that the light curve shows a stable orbital waveform on timescales of days, but alters slowly over timescales on the order of years.

% Spectroscopic characterisation
As well as time-series photometry, time-resolved spectra have also been obtained for \arsco. The spectra in the visible show a blue continuum added to the red spectrum of an M5-type dwarf, with strong Balmer and neutral helium lines in emission \citep{Marsh2016}. The radial velocity variation of the emission lines suggests that they originate near the surface of the M-dwarf facing the white dwarf. \citet{Garnavich2019} found the Balmer lines to show significant structure as they vary over the orbital period. Similar to what is observed for the orbital modulation of the light curve, the Balmer lines show peak flux between phases 0.4 and 0.5, and minima around phase 0.1\footnote{The convention is that zero orbital phase is at superior conjunction of the white dwarf, when the M-dwarf is closest to us.}. The continuum shows a similar behaviour, but is further affected by fast variations on the beat period. Doppler tomography shows that most of the H$\alpha$ emission originates on the face of the secondary star, but satellite emission features are seen, which in other systems have been attributed to long-lived prominences on the secondary star \citep[e.g.][]{Parsons2016}. \citet{Garnavich2019} argued that the stability of these prominences requires the M-dwarf to have a magnetic field of the order of 100--500~G \citep{Ferreira2000}, which is comparable to the white dwarf magnetic field strength at the M-dwarf \citep[$\sim 200$~G,][]{Takata2018}, implying that interaction between the two stars' magnetic fields is likely responsible for the energy production in \arsco.
% mass in the range 0.28~M$_{\sun} < M_M \lesssim 0.45$~M$_{\sun}$ \citep{Marsh2016} from the obtained radial velocities combined with theoretical constraints. 
% the EW of the lines changes on the beat period, but as a consequence of the changes in continuum

%Theoretical models; maybe needs a bit more detail?
A pulsar model for \arsco\ is supported by polarimetric observations, which detected strong linear polarisation (as high as 40 per cent) modulated on both the spin and beat periods \citep{Buckley2017}. The observed broad-band spectrum is associated with synchroton radiation from relativistic electrons accelerated as the white dwarf magnetosphere sweeps past the M-dwarf, resembling the emission from neutron star pulsars, but with a different mechanism whose details remain a mystery. Most likely it relates to direct interaction between the two stars, resulting in emission from the surface or coronal loops of the M-dwarf, from the magnetosphere of the white dwarf, or possibly through an associated bow shock \citep{Geng2016, Katz2017, Takata2017, PotterBuckley2018, Plessis2022}. A collimated relativistic outflow, i.e. a jet, has also been considered, but seems to be ruled out by radio observations \citep{Marcote2017}.

% Why is AR Sco important
\arsco's nature, and the rapid spin-down of its white dwarf suggests that it could be a missing link between the two major classes of accreting magnetic white dwarfs, polars and IPs, which are distinguished by whether or not the white dwarf's spin is synchronous with the orbit of the binary \citep{Katz2017}, but how AR~Sco attained its current configuration is unknown. A prior stage of accretion-driven spin-up is required to explain the current spin period. However, the high magnetic field estimated for the white dwarf in \arsco\ given its measured spin-down \citep[100--500~MG,][]{Katz2017, Buckley2017} presents a significant challenge, as an extremely high mass transfer rate would have been required in the past to crush the magnetosphere down to a co-rotation radius matching the 2~minute spin-period. In an alternative model, \citet{Lyutikov2020} argued the magnetic field of \arsco\ must be only $\sim 10$~MG to allow its spin-up, and that this is consistent with the observed spin-down rate if the material leaving the M-dwarf is originally neutral. In that case, the material can approach the white dwarf without interacting, until it is ionised by the white dwarf itself. This model requires both that the white dwarf temperature be such that the material is ionised only near the white dwarf ($T_\mathrm{eff} \approx 12\,000$~K), and that the heated face of the M-dwarf is cool enough not to ionise the gas before it heads towards the white dwarf. Although the first condition could be fulfilled \citep{Garnavich2021}, the heated face of the M-dwarf seems to be hot enough that the material would have a high ionisation rate \citep{Garnavich2019, Garnavich2021}.

Another potential solution that reconciles the need for accretion-driven spin-up and the apparently high magnetic field was put forward by \citet{Schreiber2021}. They proposed that \arsco\ only became magnetic as a result of a crystallisation- and rotation-driven dynamo. In this model, \arsco\ would originally be non-magnetic, allowing for straightforward accretion-driven spin-up. When crystallisation starts to occur in the core of the cooling white dwarf, strong density stratification combined with convection creates the conditions for a dynamo, generating the magnetic field \citep{Isern2017}. If the field is strong enough, the rapid transfer of spin angular momentum into the orbit may cause the binary to detach and mass transfer to cease, leading to a system such as \arsco. %The transfer of spin angular momentum to the orbit could thus partially counteract the effects of magnetic breaking and gravitational wave radiation, leading to larger orbital period derivatives than predicted by \citet{Knigge2011}. 

% What we do here
The many open issues surrounding the modelling of \arsco, as well as its potential central role for constraining the evolution of CVs, make precise observational constraints a valuable asset. In this work, we analyse high-speed photometry of \arsco\ obtained with ULTRACAM \citep{ultracam}, ULTRASPEC \citep{ultraspec}, and HiPERCAM \citep{Dhillon2021} spanning seven years (2015--2022), and time-resolved spectra obtained with the Intermediate-dispersion Spectrograph and Imaging System\footnote{https://www.ing.iac.es/Astronomy/instruments/isis/} (ISIS) at the William Herschel Telescope (WHT) and X-shooter at the Very Large Telescope (VLT). We further constrain the spin-down of the white dwarf using two different techniques, pulse-time arrival modelling and Fourier analysis. We probe for evidence of orbital period change by modelling spin and beat frequencies independently in the Fourier analysis. Based on our photometry, we discuss the possibility of using \arsco\ as a cosmic clock for timing observations. Using our spectroscopic data, we perform modulation Doppler tomography \citep{Steeghs2003}, extending the work of \citet{Garnavich2019}. We also measure the rotational velocity of the M-dwarf as function of orbital phase to probe for variability that could confirm that the M-dwarf is Roche lobe filling.

\section{Observations}

\subsection{Photometry}
\label{sec:phot}

We started a monitoring campaign of \arsco\ shortly after its discovery in 2015. We used three different high-speed photometers: ULTRACAM at the 3.5-m ESO New Technology Telescope (NTT), ULTRASPEC mounted at the 2.4-m Thai National Telescope (TNT), and HiPERCAM at the 10.4-m Gran Telescopio Canarias (GTC) \citep{ultracam, ultraspec, Dhillon2021}. We observed it on a total of 68 different nights, over a time span of almost seven years (2015--2022). Our cadence varied between 0.1 and 9.1 seconds, with an average of 3.3~s. This includes the readout time, which varied with the configuration and readout mode, but was never longer than 15~ms. With ULTRASPEC, we typically used a $g$ filter, with the exception of three nights when a red-blocking KG5 filter (effectively $u+g+r$) or an $r$ filter was installed. For ULTRACAM and HiPERCAM, dichroic beam splitters allow simultaneous observations in three or five different filters, respectively. Filters $u,g,r$ or $u,g,i$ were used for ULTRACAM until 14 May 2017, after which they were replaced with similar filters whose cut-on/off wavelengths match those of the typical Sloan Digital Sky Survey (SDSS) filters, but with a higher throughput, being sometimes referred to as $u_s, g_s, r_s, i_s, z_s$ or "super" SDSS. These "super" filters were also used for the HiPERCAM observations. For simplicity, we omit the $s$ subscript in the text, but details of each of our runs and filters used can be found in Table~\ref{tab:observations}.

All datasets were reduced with the dedicated HiPERCAM data reduction pipeline\footnote{https://github.com/HiPERCAM/hipercam}. We performed bias subtraction, and flat field correction using skyflats taken during twilight. We also used the pipeline to carry out differential aperture photometry with a variable aperture size, set to scale with the seeing estimated from a point-spread function (PSF) fit. We used the same comparison star for all observations, with the exception of our two HiPERCAM runs, for which the chosen comparison star was not in the field of view. Our default comparison star was {\it Gaia} EDR3 6050299715251225344 ($G = 13.3$, $G_{BP} - G_{RP} = 2.3$), whereas for HiPERCAM we used {\it Gaia} EDR3 6050298302203861888 ($G = 15.2$, $G_{BP} - G_{RP} = 1.9$). For reference, \arsco\ itself has reported values of $G =15.0$, and $G_{BP} - G_{RP} = 1.4$.

To extend our baseline to earlier years, we retrieved data for \arsco\ from the Catalina Real-Time Transient Survey \citep[CRTS,][]{Drake2009}. These data span from 2005 August 01 to 2013 July 08. The typical CRTS exposure time of 30~s is short enough that the spin and beat periods are not completely smeared out.

The observing times of all datasets were corrected to Barycentric Julian Date (BJD) in the Barycentric Dynamical Time (TDB) reference system.

\subsection{Spectroscopy}

We obtained spectroscopic observations of \arsco\ in 2016 both with X-shooter at the 8.2~m VLT \citep{Vernet2011}, and with ISIS at the 4.2~m WHT.

\begin{figure}
	\includegraphics[width=\columnwidth]{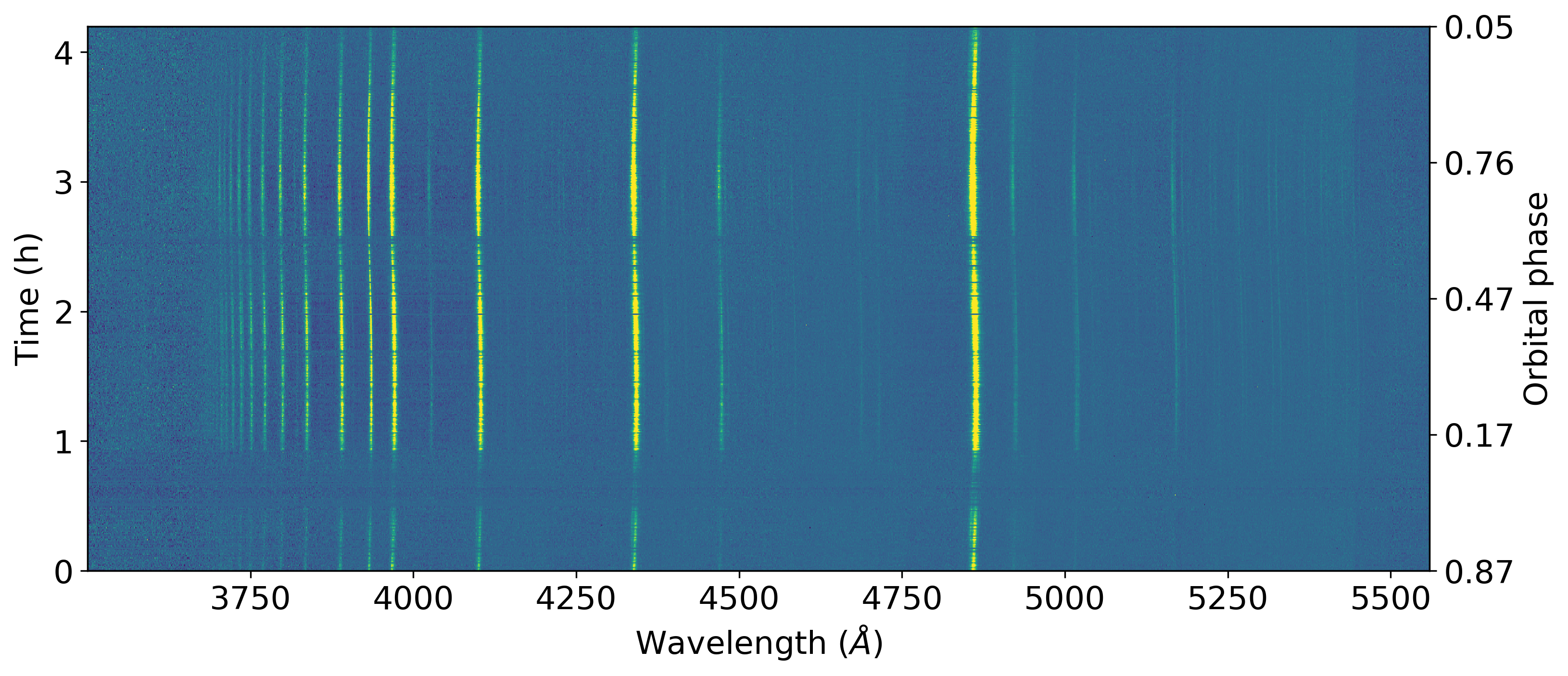}
    \caption{Trailed spectra taken with the X-shooter UVB arm. The spectra have been normalised by the continuum prior to plotting. Yellow indicates stronger flux. Many emission lines, which generally trace the heated face of the M-dwarf, can be seen.
    }
    \label{fig:uvb_trail}
\end{figure}

The X-shooter observations were carried out using a 1~arcsec slit for the UVB arm (300-559.5~nm, $R = 5400$), and 0.9~arcsec slits for the VIS (559.5-1024~nm, $R = 8900$) and NIR (1024-2480~nm, $R= 5600$) arms. In the UVB arm, we limited our exposure time to 16.1~s to allow for sampling of the beat/spin variability and obtained 689 exposures (readout time was 14~s). A fraction of the spectra that were taken continuously (the observations were briefly interrupted by clouds) is shown in Fig.~\ref{fig:uvb_trail}. For the VIS arm, we obtained 42 exposures of 354~s ($\simeq 3$ times the beat period; readout time was 58~s). The same exposure time was used for the NIR arm, but the difference in the way readout is carried out allowed 47 frames to be taken (with a readout time of 8.2~s). The spectra were reduced using the standard procedures within the ESO {\sc reflex} reduction pipeline \footnote{http://www.eso.org/sci/software/reflex/}, including telluric line removal using {\sc molecfit} \citep{molecfit1,molecfit2}.

\begin{figure}
	\includegraphics[width=\columnwidth]{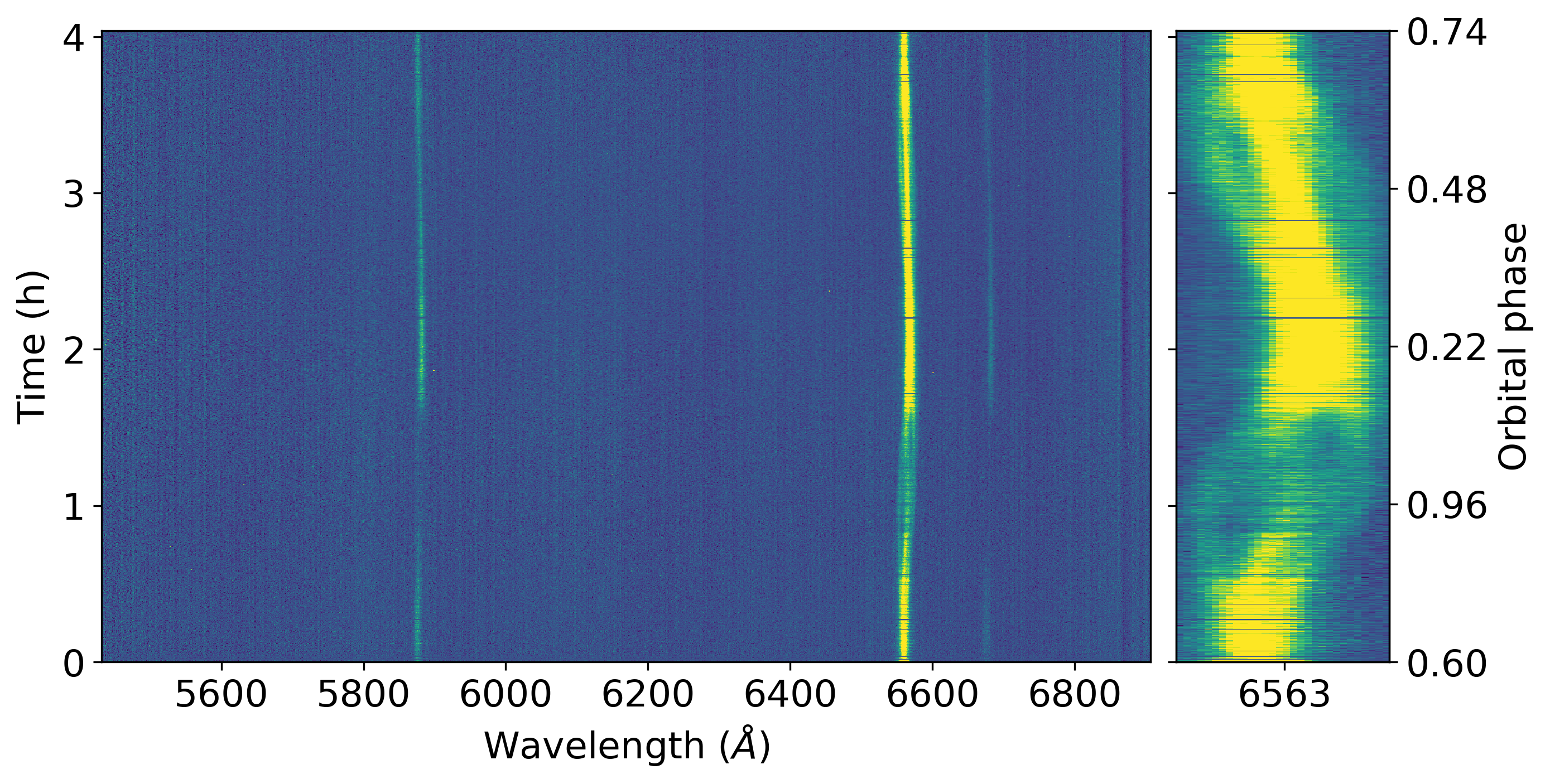}
    \caption{Trailed spectra taken with the ISIS red arm. The spectra have been normalised by the continuum prior to plotting. The panel to the right zooms into H$\alpha$, which shows some substructures that will be discussed in Section~\ref{sec:dopper}.
    }
    \label{fig:red_trail}
\end{figure}

ISIS observations were obtained using the R600B and R600R gratings in the blue and red arms, respectively. The central wavelength was set to 4500~\AA\ for the blue arm, and to 6200~\AA\ for the red arm. We used a 1.0~arcsec slit and set the exposure time to 7.3~s (blue) and 8.0~s (red), whereas the readout time was $\approx 70$~s. 1078 and 1081 consecutive exposures were obtained for blue and red arms, respectively. All spectra were de-biased and flat-fielded using the {\sc starlink}\footnote{https://starlink.eao.hawaii.edu/starlink} packages {\sc kappa}, {\sc figaro} and {\sc convert}. Optimal spectral extraction was carried out using {\sc pamela} and wavelength calibration was performed with {\sc molly} \citep{Marsh1989}. Spectra taken with the red arm are shown as a trail in Fig.~\ref{fig:red_trail}.

\section{Data analysis}

\subsection{Pulse arrival time estimates}

We have estimated individual pulse arrival times for each of our observing runs with ULTRASPEC, ULTRACAM, or HiPERCAM. As the pulse timing depends on the filter used \citep[see][and Section~\ref{sec:clock}]{Gaibor2020}, we have only used data taken with the $g$ filter for this purpose. Our method is similar to that previously employed by \citet{Stiller2018} and \citet{Gaibor2020}, who used a Gaussian function to model the beat pulse peak. We first estimated the location of each peak using previously determined ephemeris for the beat period. Next, we cross-correlated the data around each peak with a Gaussian function with a fixed width set to minimise the residuals when fitting the ephemeris. We experimented with standard deviation ($\sigma$) values for the Gaussian ranging from 5 to 20~s, finding an optimal value of 10.1~s. To determine the maximum of the cross-correlation function, which is the estimate for the location of the peak, we determined the value at which the derivative of the cross-correlation changes sign from positive to negative using Newton-Raphson iteration. The initial ephemeris was also used to estimate the cycle number of each peak measurement. We repeated the procedure recalculating the cycle numbers using the derived ephemeris in order to get self-consistent values; this was repeated until the assumed and fitted ephemeris showed no significant change.

\begin{figure}
	\includegraphics[width=\columnwidth]{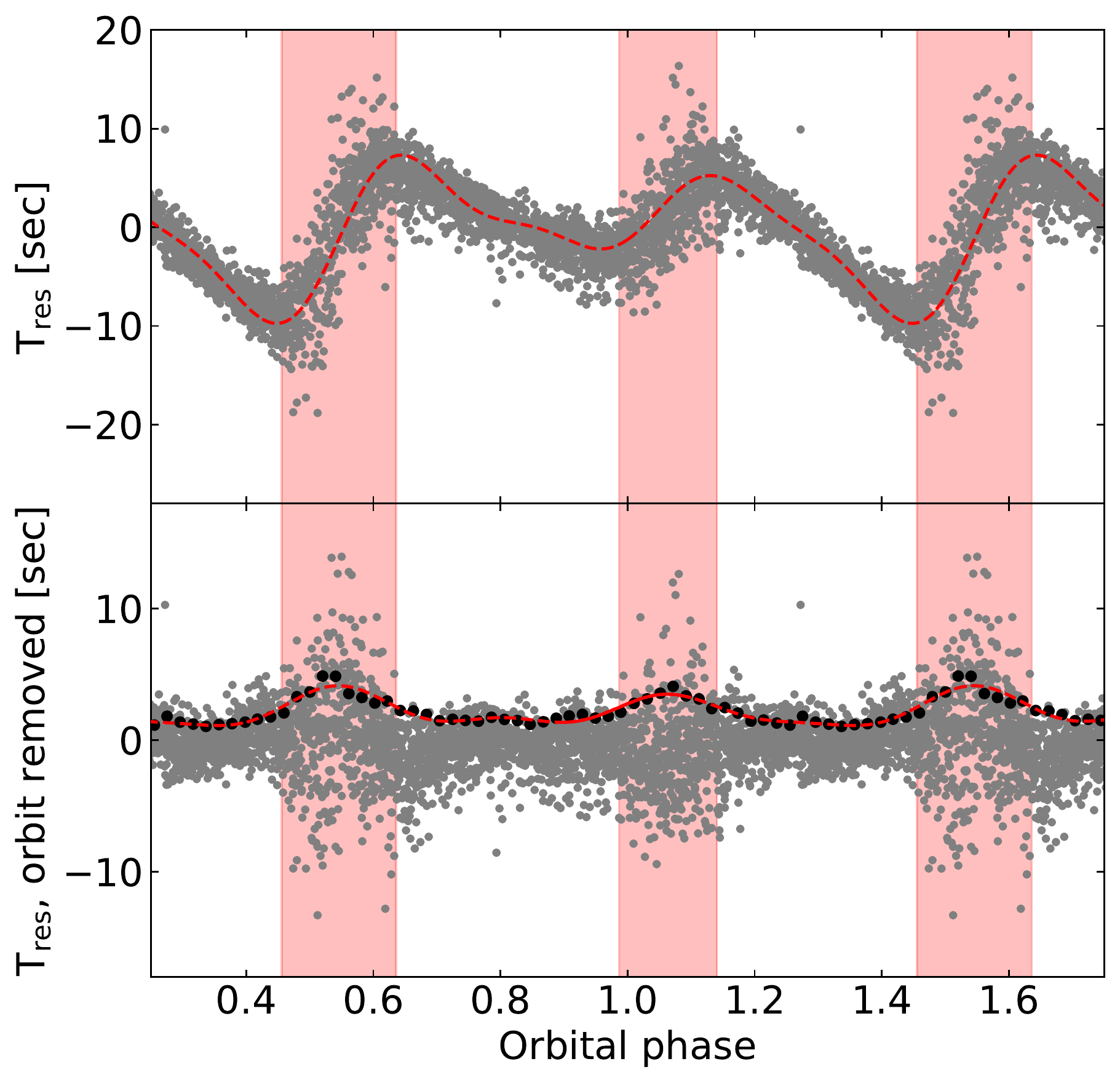}
    \caption{The grey dots in the top panel show the residuals as a function of orbital phase when fitting the pulse arrival times with only a second order polynomial. The red dashed line is a Fourier fit to the residuals, which show clear dependence on the orbital phase. The bottom panel shows the residuals when an orbital component modelled by a Fourier series is included in the fit. The black points show the root mean square binned to 50 points, fitted by a Fourier series (red dashed line). There are still phases with a large amount of scattering, marked by the regions shaded in red, which were excluded from subsequent fits.
    }
    \label{fig:orbitdiffs}
\end{figure}

Data points affected by clouds were manually flagged in each run, and for each peak we computed the resulting number of bad points. Peak measurements were only accepted if the number of good points exceeded the number of bad points by at least a factor of three. Peaks were also rejected if the difference between the measurement and predicted value based on the approximate ephemeris was more than 10 per cent of the beat period. This method resulted in 3195 good pulse measurements (table included as Supplementary Material). However, the beat pulses are known to show an orbital phase dependence \citep{Stiller2018}, which needs to be corrected for. To account for this orbital modulation, we performed an initial fit to the derived pulse times, and calculated the observed minus calculated values (O-C) as a function of orbital phase $\phi_\mathrm{orb}$. We then modelled the O-C behaviour with a Fourier series, and subtracted the modelled O-C orbital behaviour from each obtained pulse arrival time\footnote{We have used a Fourier series with four terms in the form $S_i\sin(2\pi i \phi_\mathrm{orb}) + C_i\cos(2\pi i \phi_\mathrm{orb})$ with ($S_i, C_i$) coefficients in seconds given by (-1.752, 2.165), (4.723, -2.769), (-0.929, 0.734), (0.541, -1.382).} (see Fig~\ref{fig:orbitdiffs}). Note that the values supplied as Supplementary Material do not include this correction, as combining them with more data can change the required number of Fourier components and their coefficients.

This same procedure cannot be applied to the CRTS data, since the pulses are not individually resolved at the CRTS cadence. We have instead determined the maxima by fitting a cosine to the data. We first phase-folded the data to the orbital ephemeris of \citet{Marsh2016}, and modelled the orbital variability with a Fourier series. The orbital contribution was then subtracted from each measurement. Given the long-time span of the CRTS observations, the spin-down can have a measurable effect in the data. To account for this, we subtracted a quadratic term obtained from fitting the ULTRASPEC/ULTRACAM/HiPERCAM data from the measured times. We then fitted cosines with a period fixed to the beat period to the CRTS data divided into two datasets at the midpoint of observations (2008 September 17). Data taken at phases showing excess scatter (see next paragraph) were excluded. This resulted in two measurements for the maximum of the beat pulse extending our baseline back to 2005, with each representing the mean pulse time from three years of CRTS coverage.

\begin{figure}
	\includegraphics[width=\columnwidth]{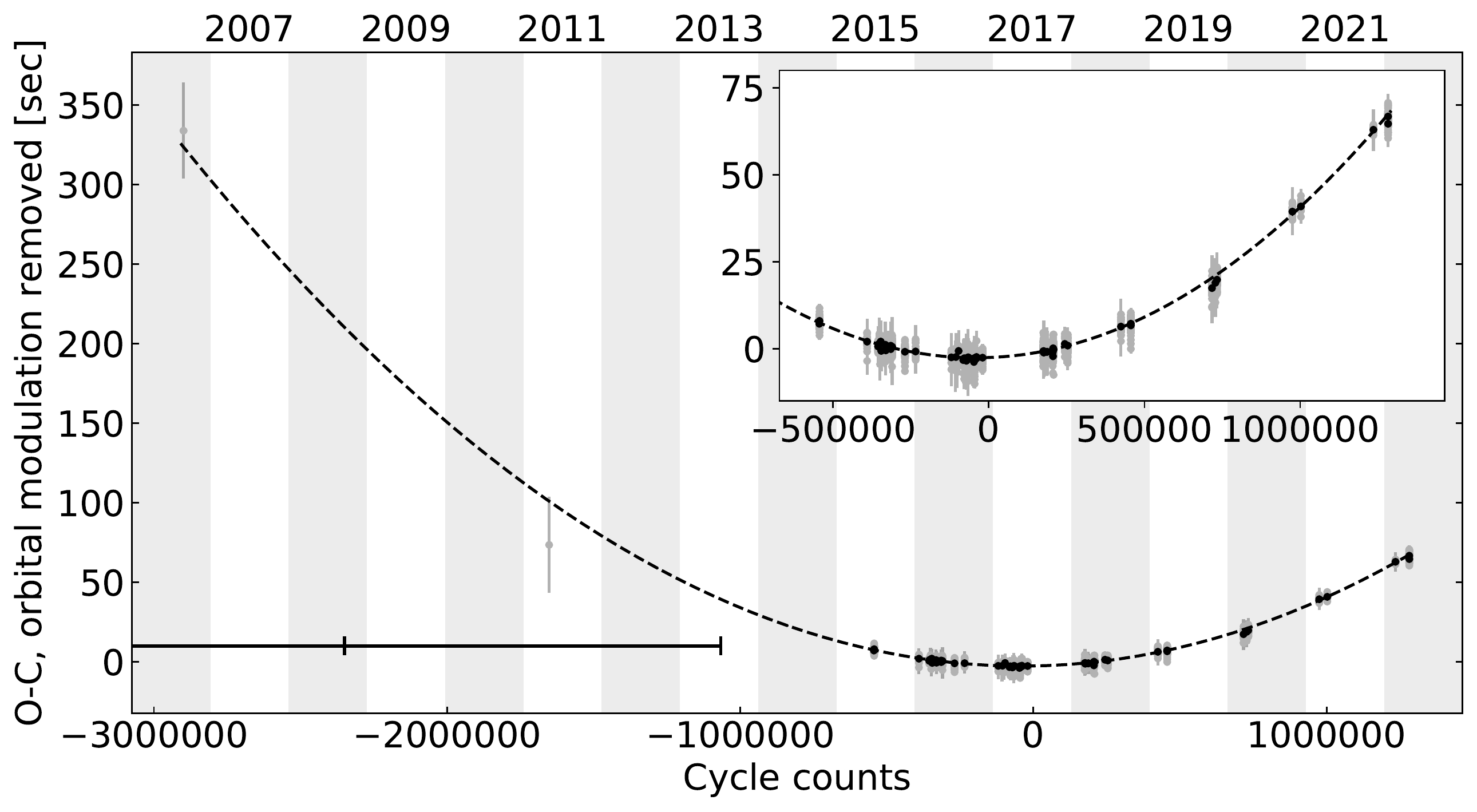}
    \caption{The $y$-axis shows the residuals (observed minus calculated, O-C) after the linear part of the ephemeris is subtracted, and the orbital modulation is corrected. The alternate grey/white columns mark different years, indicated at the top. The two left-most points correspond to the estimates from the CRTS data; the black bar at the bottom indicates the time coverage of these measurements (more details in the text). Measurements from 2015 onward correspond to ULTRASPEC/ULTRACAM/HiPERCAM data, and are shown in more detail in the inset. Individual measurements are shown in grey, and black points show the data binned every 20 points. The dashed line is our quadratic fit.}
    \label{fig:ominusc}
\end{figure}

Finally, we carried out a least-squares fit of a quadratic function to the measured times and estimated cycle numbers. We excluded from our fit the measurements taken around orbital phases 0.05 and 0.55 (more specifically, in the ranges 0.987--0.140 and 0.455 and 0.635), which show excess scatter because the maxima are less well defined around these phases \citep[see Fig~\ref{fig:orbitdiffs}, as well as ][fig. 2]{Gaibor2020}. These orbital phase limits were set to exclude points with a root-mean-square deviation (RMS) larger than 2.5~s after correcting for orbital effects. For the remaining 2135 measurements, the uncertainty was recalculated as the RMS at the given orbital phase. Fitting these measurements, we obtained the ephemeris:
\begin{eqnarray}
    T_\mathrm{beat~max} (BJD) = 2457941.6688819(20) \nonumber+\\ 0.001368045813(8) E \nonumber +\\ 4.53(8) \times 10^{-16} E^2,
\end{eqnarray}
where $E$ is the integer cycle number. The quoted one-sigma uncertainties were determined via bootstrapping. The quadratic coefficient is equivalent to $-0.5 \dot{\nu}_\mathrm{beat} P_\mathrm{beat}^3$, thus implying a beat frequency derivative of $-4.74\pm0.08 \times 10^{-17}$~\hzs, consistent with the estimates of both \citet{Stiller2018} ($-5.14\pm0.32 \times 10^{-17}$~\hzs) and \citet{Gaibor2020} ($-4.82\pm0.18 \times 10^{-17}$~\hzs) within $1~\sigma$, and with a 50-$\sigma$ significance. Figure~\ref{fig:ominusc} shows the fit of the quadratic component of the ephemeris.

\subsection{Fourier Analysis}

In addition to using our photometric data to perform pulse arrival time estimates, we also carried out a Fourier analysis. As in the previous section, we focused on observations taken with the $g$ filter. To minimise systematic effects, we excluded any data points affected by clouds, and only utilised data employing our default comparison star.

\begin{figure}
	\includegraphics[width=\columnwidth]{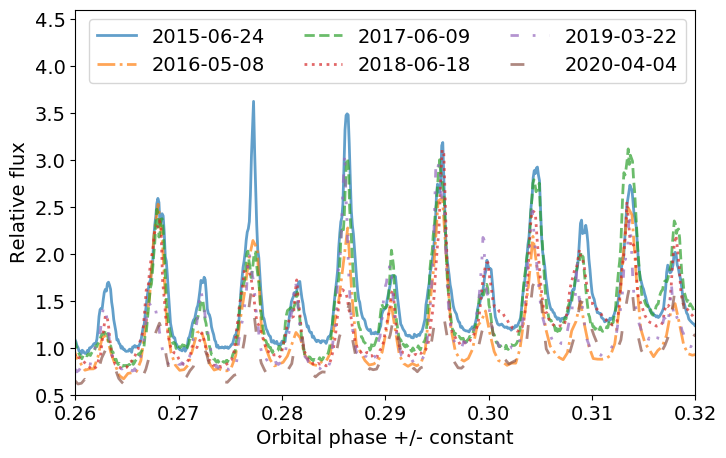}
    \caption{Sample runs from different years. The orbital phase is accurate for the 2015 run, but we have slightly displaced the runs from other years to align the pulses, in order to illustrate their changing amplitude.}
    \label{fig:some_pulses}
\end{figure}

To identify the main harmonics and combinations contributing to the observed waveform, we calculated the Fourier transform of the data, simultaneously modelled the identified dominant contributions with a Fourier series, subtracted the model from the light curve, and then re-calculated the Fourier transform of the residuals. The modelling was done by performing a least-squares fit to the data with orbital contributions modelled as sine waves, whereas the spin and beat contributions were modelled as cosines (we differentiate between sines and cosines to make it clear what phase 0 implies). Timing parameters were common to all data, whereas amplitudes and phases were fitted individually to each run. We started by subtracting the fundamental frequency and the first harmonic for the orbital ($\Omega$), spin ($\omega$), and beat ($\omega-\Omega$) frequencies. This revealed the next main contribution to be the second harmonic of the beat frequency $2(\omega-\Omega)$, followed by the combinations $2\omega - \Omega$, $4\omega - 2\Omega$, $\omega - 2\Omega$, the third harmonic of the beat frequency 4($\omega-\Omega$), $\omega + \Omega$, and $2\omega - 3\Omega$. At this point, the dominant peak became again the beat period, which is not perfectly modelled using this procedure due to the varying amplitude of the peak \citep[see][and Fig.~\ref{fig:some_pulses}]{Stiller2018}. Figure~\ref{fig:ft_prew} illustrates this pre-whitening process.

\begin{figure}
	\includegraphics[width=\columnwidth]{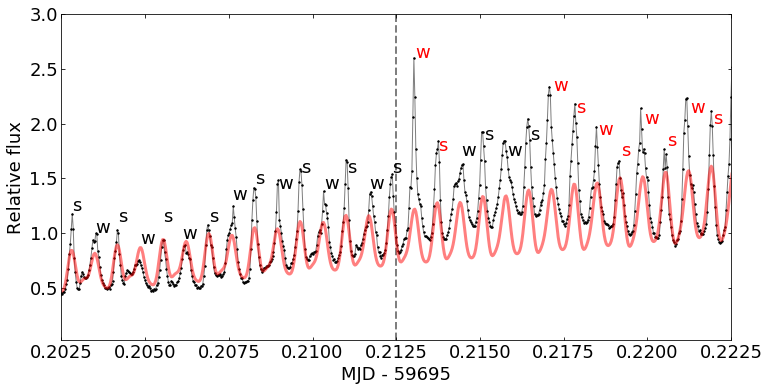}
    \caption{The inversion of pulse strength observed in our last run. The data are shown as black dots, with our Fourier model shown in red. Data beyond the grey dashed line was excluded from the Fourier fit. The typical behaviour of AR~Sco consists of alternating strong (s) and weak (w) pulses, whose intensity can get very similar in particular for certain orbital phases. However, after MJD = 59695.2125 there was a particularly strong pulse where a weak pulse was expected. After some pulses with the normal behaviour, the inversion becomes persistent, which caused a large increase in $\chi^2$.}
    \label{fig:bad_pulses}
\end{figure}

\begin{figure*}
	\includegraphics[width=0.8\textwidth]{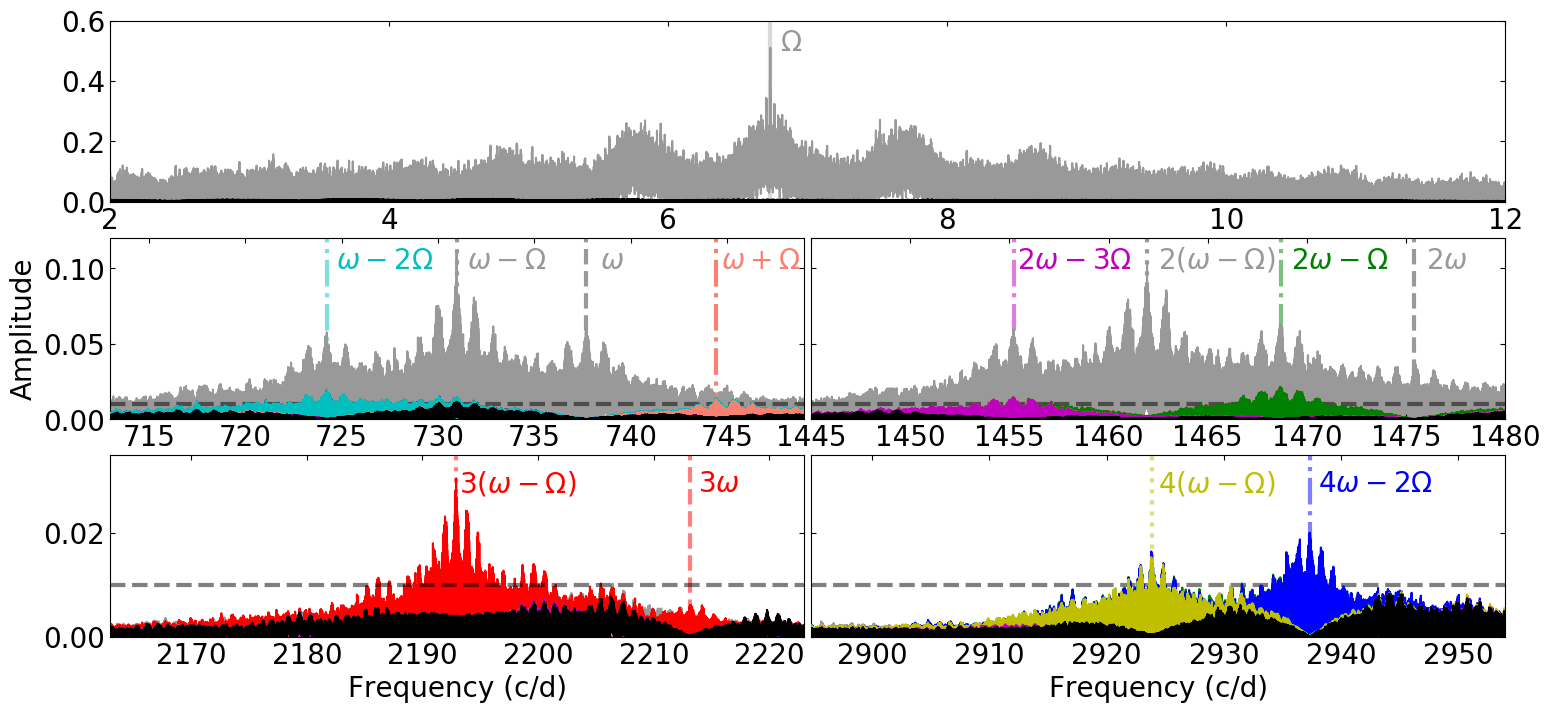}
    \caption{Fourier transforms around the orbital frequency (top panel), spin and beat fundamental frequency (middle-left), first (middle-right), second (bottom-left), and third (bottom-right) harmonics. The original Fourier transform is shown in grey, with the peaks that were initially subtracted indicated in the same colour as vertical lines. Each subsequent iteration of our pre-whitening procedure is shown in the same manner, i.e. the Fourier transform and the peaks subtracted before the next iteration in the same colour. In black we show the Fourier transform after our final model is subtracted. The black dashed line is placed at an amplitude of 0.01 (1 per cent) to facilitate comparison between the middle and bottom panels. Though some combinations remain visible in the amplitude spectrum, the main contribution is again the beat frequency. Given the sparse sampling over many years, many aliases are seen, in particular one-day aliases at this scale.}
    \label{fig:ft_prew}
\end{figure*}

Once we identified the main contributions to the observed light curve, we carried out a fit of the light curve including all of these contributions using Markov-Chain Monte-Carlo (MCMC) implemented with {\sc emcee} \citep{emcee}. We initially performed a fit with amplitudes and phases of all contributions allowed to vary freely, but each set to be constant throughout the whole span of our data. This led to a poor fit of the light curve, as the amplitudes of the peaks are not consistent over time (as shown in Fig.~\ref{fig:some_pulses}). We then opted for initially fitting each of our runs individually, with the timing parameters fixed to the values obtained from a least-squares fit to all of the data, but amplitudes and phases free. The obtained amplitudes and phases were then fixed, and we carried out the MCMC fit with only the timing parameters, $T_{0}^{\mathrm{orb}}, T_{0}^{\mathrm{beat}},  \nu_\mathrm{beat}, \nu_\mathrm{spin}, \dot{\nu}_\mathrm{beat}, \dot{\nu}_\mathrm{spin}$, as free. Since our runs typically do not cover a full orbit, the orbital period itself is not well constrained from sinusoidal fits to our data, but it can be indirectly obtained by modelling the beat and spin frequencies separately. We therefore allowed the spin and beat periods to vary freely, and calculated the orbital period from their difference. The linear term corresponding to $T_0$ was allowed to be free for the orbital and beat periods, but constrained to be within one period of our initial guess. For the spin period, as well as spin/orbit combinations, $T_0$ was set to be equal to the beat $T_0$, but a phase shift was allowed. A phase shift was also allowed for all harmonics. In addition to the periods themselves, we also included the period derivative as a free parameter for both spin and beat in our model. We have not constrained them to be equal in order to probe for a significant orbital period derivative. Data points whose residuals were more than three standard deviations away from an initial least-squares fit were excluded from the MCMC fit to avoid occasional strong peaks, which typically have low uncertainties and could bias our fit. Uncertainties were inflated by a factor of 8 to achieve reduced $\chi^2 \approx 1$ (though we report the $\chi^2$ prior to this renormalisation below). Additionally, we excluded data taken towards the end of our last run, 2022 April 26 (MJD > 59695.2125), because we noticed that the inclusion of these data led to an increase in $\chi^2$ of at least a factor of two, leading to a poor solution. The reason behind this seems to be an inversion in the strength of consecutive peaks (see Fig.~\ref{fig:bad_pulses}), possibly caused by a change in which pole generates the strongest synchrotron emission power. This is also eventually observed in other runs, but has a particularly damaging effect to the fit when occurring in the last portion of data, which plays an important role in constraining the frequency derivatives. With the current data we see no evidence for a periodical behaviour of such inversion, which seems to be stochastic. Future observations will reveal whether this behaviour persists after our last run.

\begin{figure*}
	\includegraphics[width=0.8\textwidth]{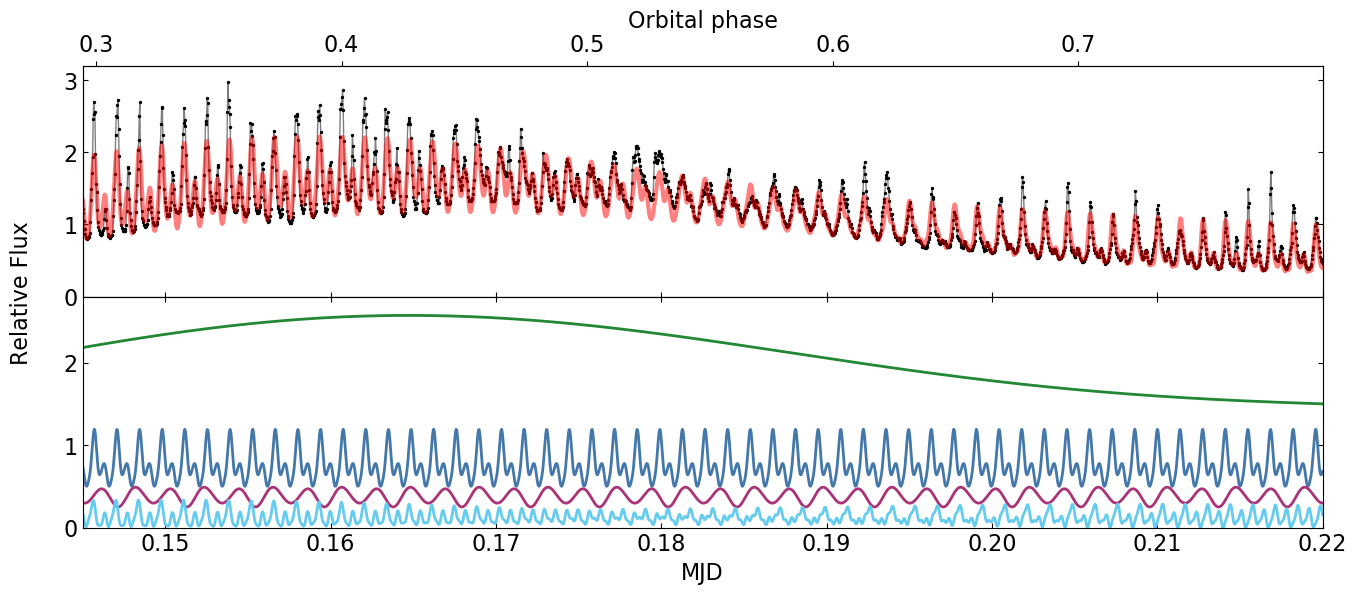}
    \caption{The observed data are shown in the top panel as black dots connected by grey lines. The light curve model is shown in red. In the bottom panel, we illustrate the different contributions, which are from top to bottom: orbital ($\Omega$ and $2\Omega$, in green), beat ($\omega - \Omega$, $2\omega - 2\Omega$, $3\omega - 3\Omega$, and $4\omega - 4\Omega$, in blue), spin ($\omega$, $2\omega$ and $3\omega$, in purple), and other combinations ($2\omega - \Omega$, $4\omega - 2\Omega$, $\omega - 2\Omega$, $2\omega - 3\Omega$, and $\omega + \Omega$, in cyan).}
    \label{fig:model_lc}
\end{figure*}

When running the MCMC fit, we at first applied a prior forcing the difference between spin and beat derivatives to be consistent with the upper limit determined by \citet{Peterson2019}, that is $\dot{\nu}_\mathrm{spin} - \dot{\nu}_\mathrm{beat} \lesssim 3.8 \times 10^{-20}$~\hzs. We obtained a solution with $\chi^2 = 18415210$ (for 153527 datapoints), whose obtained parameters are given in Table~\ref{tab:fourier_fit}. Next we removed the prior and let $\dot{\nu}_\mathrm{spin}$ and $\dot{\nu}_\mathrm{beat}$ vary freely. Without the prior, we obtain a solution with a lower $\chi^2$ of 18381704 for the same number of points. Figure~\ref{fig:model_lc} shows a section of the data with this best-fit model, and Figure~\ref{fig:corner} shows a corner plot of the resulting timing parameters. The obtained beat period and spin and beat frequency derivatives are consistent at the 99 per cent confidence level for the two solutions, though the value of the spin period and of $T_{0}^{\mathrm{beat}}$ shift significantly. Notably, in both cases we obtain $\dot{\nu}_\mathrm{beat}$ consistent with our pulse time arrival analysis. The main distinction between the two fits is that, without a prior, the difference between our spin and beat period derivatives implies a significant $\dot{\nu}_\mathrm{orbit} = (8.43\pm0.37) \times 10^{-19}$~\hzs, higher than the upper limit suggested by \citet{Peterson2019}, whereas the value is naturally lower, $\dot{\nu}_\mathrm{orbit} = (3.68\pm0.17) \times 10^{-20}$~\hzs, when the upper-limit prior is applied. We further discuss this result in Section~\ref{sec:nudots}.
%The value of $T_{0}^{\mathrm{orb}}$ differs from that determined from radial velocities given the offset between the maximum/minimum flux and phases 0/0.5. 

\begin{table*}
	\centering
	\caption{The two solutions obtained from the Fourier fit to the light curve. Uncertainties are given by the standard deviation obtained in the MCMC run.}
	\label{tab:fourier_fit}
	\begin{tabular}{ccc} % four columns, alignment for each
		\hline
		Parameter & $\dot{\nu}_\mathrm{spin} - \dot{\nu}_\mathrm{beat} \lesssim 3.8 \times 10^{-20}$~\hzs & $\dot{\nu}_\mathrm{spin}$, $\dot{\nu}_\mathrm{beat}$ free\\
		\hline
		$T_{0}^{\mathrm{orb}}$ & 2457264.624698(14) & 2457264.624503(17) \\
		$T_{0}^{\mathrm{beat}}$ & 2457941.6689500836(15) & 2457941.6689500569(15) \\
		$P_\mathrm{beat}$ (days) & 0.00136804583419(44) & 0.00136804583230(45) \\
		$P_\mathrm{spin}$ (days) & 0.00135556080771(47) & 0.00135556081486(57) \\
		$\dot{\nu}_\mathrm{beat}$ ($10^{-17}$~Hz/s) & $-4.945\pm0.007$ & $-4.963\pm0.006$ \\
		$\dot{\nu}_\mathrm{spin}$ ($10^{-17}$~Hz/s) & $-4.941\pm0.006$ & $-4.878\pm0.007$ \\
		Reduced $\chi^2$ & 119.95 & 119.73 \\
		\hline
	\end{tabular}
\end{table*}

%\begin{figure*}
%	\includegraphics[width=\textwidth]{full_ft_all.pdf}
%    \caption{{\bf [PLACE HOLDER, waiting for final fit to add pre-whitened FTs]} Fourier transform for individual years, and for the whole dataset.}
%    \label{fig:ft}
%\end{figure*}

\subsection{Doppler tomography}
\label{sec:dopper}

\begin{figure*}
	\includegraphics[width=\textwidth]{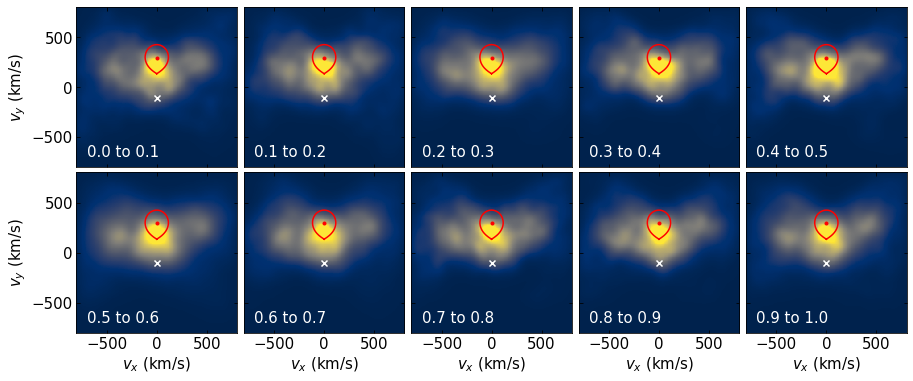}
    \caption{Doppler tomography of the H$\alpha$ line for different intervals of beat phase, indicated in the bottom left corner of each panel. The white cross marks the position of the white dwarf, the red line indicates the Roche lobe of the companion, and the red dot its centre of mass. The side lobes seen at all phases have been previously reported by \citet{Garnavich2019}.}
    \label{fig:dop_red}
\end{figure*}

\begin{figure}
	\includegraphics[width=\columnwidth]{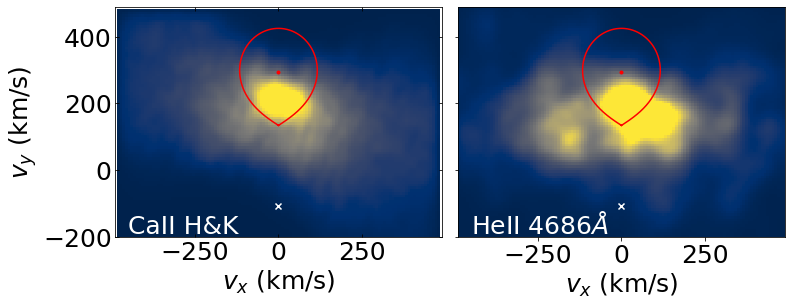}
    \caption{Doppler tomography for the CaII H\&K lines (left panel), and for HeII 4686~\AA\ (right panel). Both emissions originate mainly from the irradiated Roche lobe of the M-dwarf, though the HeII line shows an excess near the trailing face of the star, as also seen by \citet{Garnavich2019}.}
    \label{fig:dop_caii_heii}
\end{figure}

Similarly to \citet{Garnavich2019}, we have applied the Doppler tomography technique to our obtained spectra. This technique consists of mapping the observed line profiles at different orbital phases into velocity space, thereby allowing the line emission distribution in the binary to be mapped \citep[see][for a review]{Marsh2001}. We started by analysing the H$\alpha$ line using our WHT spectra. \citet{Garnavich2019} previously reported the existence of long-lived prominences (see their fig. 7), whose existence, they argued, requires the M-dwarf to have a strong magnetic field ($100-500$~G). Owing to the short integration time of our exposures, we were able to search for variations of these prominences with beat period, which could indicate that they are modulated by the interaction between the magnetic field of the white dwarf and the companion, since the beat phase tracks the orientation of the white dwarf with respect to the binary orbit. As can be seen in Figure~\ref{fig:dop_red}, these features seem to persist regardless of beat phase, suggesting that they are indeed inherent to the M-dwarf. In addition, they seem to appear only for the H$\alpha$ line, whereas the emission for other lines is mostly concentrated on the face of the M-dwarf facing the white dwarf, though there are some substructures that vary from line to line (see Figure~\ref{fig:dop_caii_heii}). The detection of He~{\sc ii} is particularly important, as it suggests a high temperature on the M-dwarf's heated face that would ionise material before it leaves the star, which invalidates \citet{Lyutikov2020}'s model \citep[as previously pointed out by][]{Garnavich2021}. The location of this stream also suggests interaction close to the M-star, in agreement with the model suggest by \citet{Katz2017}, rather than to the white dwarf, as proposed by \citet{Lyutikov2020}.

%\citet{Garnavich2019} said "Redshifted Balmer-emission flashes are correlated with the bright phases of the continuum beat pulses while blueshifted flashes appear to prefer the time of minimum in the beat light curve.", but I guess we don't see evidence for that here?

%\begin{figure*}
%	\includegraphics[width=\textwidth]{uvb.png}
%    \caption{UVB arm (H$\beta$ \& H$\gamma$)}
%    \label{fig:dop_uvb}
%\end{figure*}

It is important to note that persistent H$\alpha$ prominences have been detected in other non-accreting systems, such as QS~Vir \citep{Parsons2016} or SDSS J1021+1744, in which they have been attributed to long-lived material near the L5 point \citep{Irawati2016}. Prominences are also shown by some CVs during low states, as reported for instance for AM~Her \citep{Kafka2008} and BB~Dor \citep{Schmidtobreick2012}. Similar to what we see here, these CVs also show strong prominences mainly in H$\alpha$, with emission in other lines appearing more concentrated near the heated face of the white dwarf companion. \citet{Kafka2008} and \citet{Schmidtobreick2012} noted that the prominences do not coincide with the location of the outer Lagrangian points, L4 and L5, but \citet{Schmidtobreick2012} hypothesised that the magnetic field could alter the Roche geometry and result in equilibrium points where the prominences are observed. A magnetic field has also been suggested by \citet{Parsons2016} to explain the prominences seen in QS~Vir, and could also be responsible for the features seen in \arsco.

\begin{figure}
	\includegraphics[width=\columnwidth]{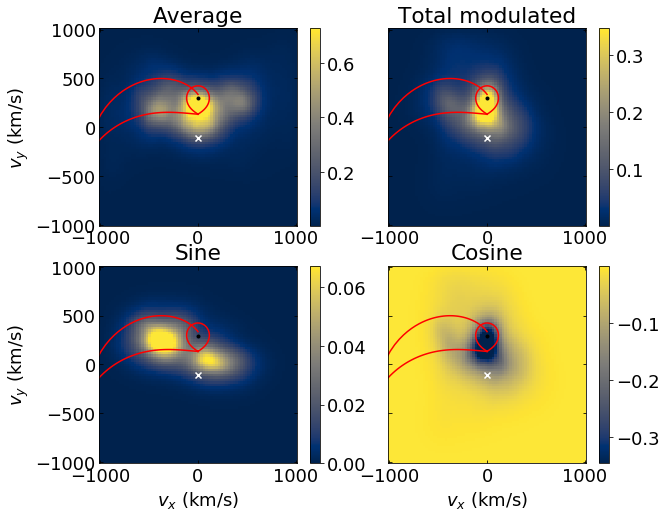}
    \caption{Modulation Doppler tomography maps of the H$\alpha$ line following \citet{Steeghs2003}. The top left panel shows the average amplitude (non-modulated), similar to Fig.~\ref{fig:dop_red}, but including all beat phases. The top right panel shows the total amplitude that is modulated on the orbital period. The bottom left panel shows modulations on a sine (peaking at orbital phase 0.5), and the bottom right shows modulations on a cosine (peaking at orbital phase 0.0). The modulated amplitude shows complex orbital-phase dependent structures, as further illustrated in Fig.~\ref{fig:vmap_red}. We include here what would be the trajectory of the accretion stream, if it existed.}
    \label{fig:moddop_red}
\end{figure}

\begin{figure}
	\includegraphics[width=\columnwidth]{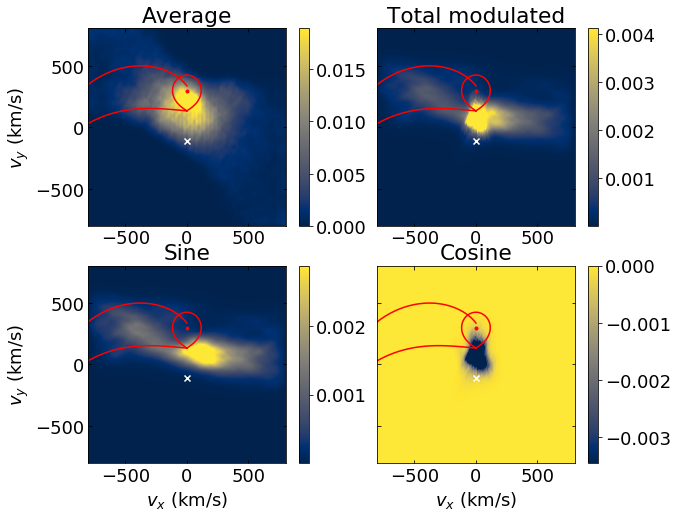}
    \caption{Same as Fig.~\ref{fig:moddop_red}, but for H$\beta$. Similar structures to those in H$\alpha$ can be seen, but with much smaller amplitude.}
    \label{fig:moddop_uvb}
\end{figure}

One of the assumptions of traditional Doppler tomography is that the flux of any point in the binary system is constant over time. However, emission that is modulated with time is often observed in interacting systems \citep[e.g.][]{Papadaki2008, Calvelo2009, Somero2012}. \citet{Steeghs2003} extended the Doppler tomography method to allow for emission modulated on the orbital period. We applied this modulated Doppler tomography to the WHT spectra H$\alpha$ line (Figure~\ref{fig:moddop_red}), and to the H$\beta$ (Figure~\ref{fig:moddop_uvb}) line using the X-shooter spectra, which have a better spectral resolution than our WHT spectra.

\begin{figure}
	\includegraphics[width=\columnwidth]{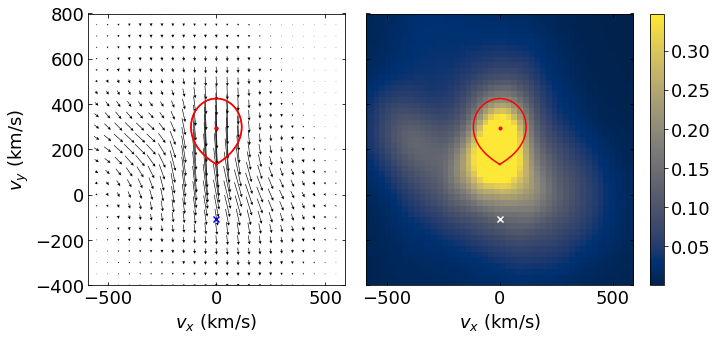}
    \caption{The left panel shows the total modulated amplitude for the H$\alpha$ line as a vector map. The direction of the arrow indicates the orbital phase, with the arrow pointing upwards for phase zero and going clockwise from there (i.e. phase 0.25 to the right, and phase 0.5 downwards, and so on). The length of the arrows is proportional to the amplitude. The panel on the right is the same as the top right panel in Fig.~\ref{fig:moddop_red}, but it is repeated here with the same scale as the vector map to facilitate comparison.}
    \label{fig:vmap_red}
\end{figure}

Both lines show components with a significant orbital period modulation. Unlike the average amplitudes, which are somewhat dissimilar for H$\alpha$ and H$\beta$ given that only the former shows strong prominences, the modulated amplitude shows a similar behaviour for both lines. The sine component, which peaks at orbital phase 0.5, shows two main components, one of which has a similar origin as one of the observed H$\alpha$ prominences, whereas the other is located near the trailing face of the M dwarf, much like seen for the HeII 4686~\AA\ line. The cosine component (peaking at phase 0.0) is mainly along the line connecting the two stars, with an extension towards the same H$\alpha$ prominence seen in the sine map. This phase-dependent behaviour is further illustrated in Figure~\ref{fig:vmap_red}. The extension seems to be located between what would be the Keplerian and stream trajectories, if there were any mass transfer between the two stars. This is similar to what is seen in, for example, U Geminorum \citep{Marsh1990}.  In \arsco's case, there is no evidence for accretion; yet, the existence of persistent prominences suggests the concentration of matter in the region where the stream would be. The modulated emission could thus result from a shock in this region resulting from interaction between the two stars' magnetic fields as the system rotates.

\subsection{Rotational velocity of the M-dwarf}

Evidence of ellipsoidal variation in \arsco's light curve suggests that the M-dwarf is filling its Roche lobe, or very nearly so \citep{Marsh2016}. For any star with a high Roche lobe filling factor, the radius along the line of sight can change significantly as it rotates, which translates into variable projected rotational velocity, $v_\mathrm{rot} \sin i$, given that $v_\mathrm{rot} = 2\pi R_{\star}/P_{\star}$ (where $R_{\star}$ and $P_{\star}$ are the radius and rotation period of the star). Our X-shooter spectra include several lines originating in the atmosphere of the M-dwarf that allow us to probe for this $v_\mathrm{rot} \sin i$ variability.

Following \citet{Parsons2018}, we selected three sets of lines for the estimates: the KI line at 7699\,\AA, the NaI lines at 8183/8193\,\AA, and the KI line at 1.252\,$\mu$m. We have used the M4.5 template from \citet{Parsons2018}, which we found to yield a better fit to these lines than the M5 template. Our approach was to find the value of $v_\mathrm{rot} \sin i$ that minimised the $\chi^2$ between the observed line and the template. We normalised and continuum-subtracted both observed spectra and template prior to fitting. We also corrected the observed spectra to the rest-frame of the M-dwarf.

\begin{figure}
	\includegraphics[width=\columnwidth]{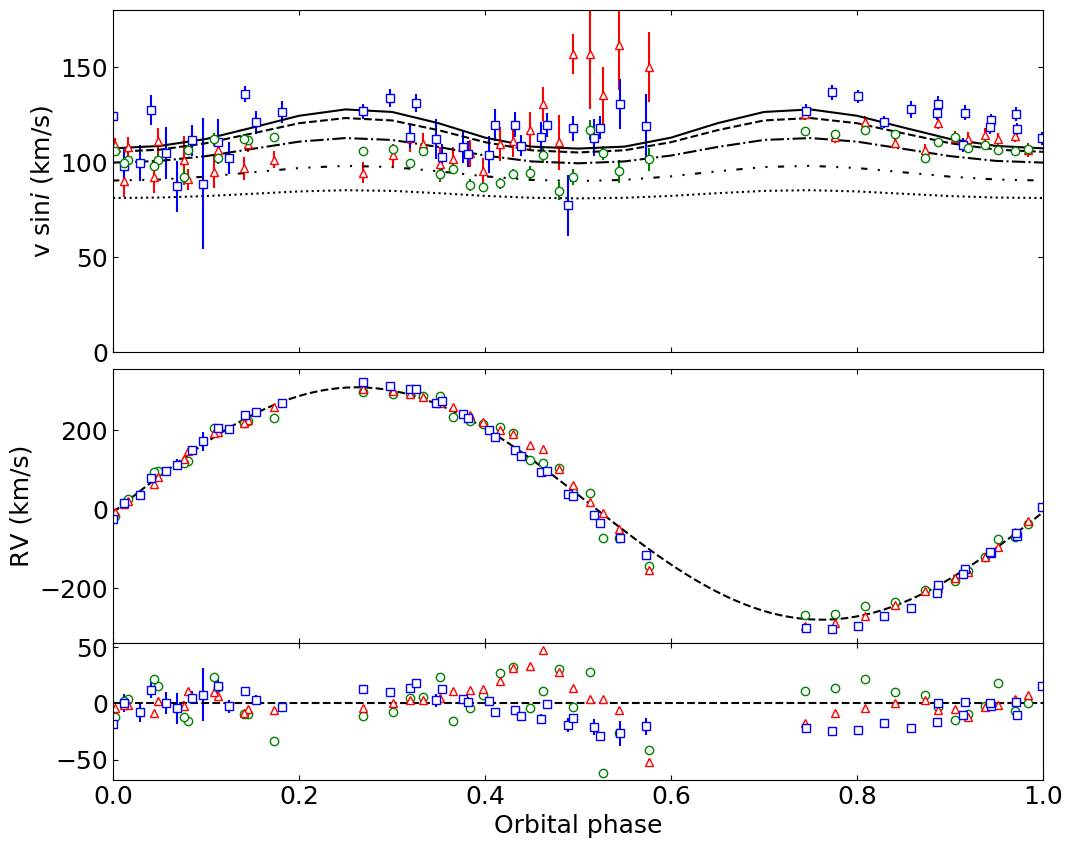}
    \caption{The top panel shows $v_\mathrm{rot} \sin i$ measurements, whereas the radial velocity is shown in the middle panel. Green circles are measurements from the NaI 8183/8193\,\AA\ line, red triangles from KI 7699\,\AA, and blue squares from KI 12522\,\AA. The orbital phase was calculated using the ephemeris of \citet{Marsh2016}. The dashed line in the radial velocity panel is a sinusoidal fit to the data, which was used to correct the wavelengths to rest frame before measuring $v_\mathrm{rot} \sin i$. The bottom panel shows residuals to this fit. The obtained parameters agree to within less than one-sigma with those reported by \citet{Marsh2016}. In the $v_\mathrm{rot} \sin i$ panel, the lines are for radii of 100, 90, 80, 70, and 60 per cent of the Roche radius (from top to bottom).}
    \label{fig:vsini}
\end{figure}

In order to obtain model values for comparison, we utilised the {\sc lprofile} code, which is part of the {\sc lcurve} package \citep{Copperwheat2010}. We calculated phase-resolved synthetic line profiles assuming the system parameters derived by \citet{Marsh2016} and an orbital inclination of 60 degrees \citep{PotterBuckley2018, Plessis2019}. The exposure length and the spectral resolution were fixed at values matching our X-shooter spectra. We estimated the full-width at half maximum (FWHM) of each line by fitting a Gaussian profile, and did the same for the 1.252~$\mu m$ line (which is the most isolated of the analysed lines). We then interpolated the observed (close to linear) relationship between FWHM and $v_\mathrm{rot} \sin i$ to estimate the $v_\mathrm{rot} \sin i$ value of each synthetic profile. We carried out this estimate both for a completely Roche-lobe filling M-dwarf, and for radii of 90, 80, 70 and 60 per cent of the Roche-lobe radius (the latter corresponds approximately to the radius of an isolated M-dwarf of mass 0.3~M$_{\sun}$). These radii are a linear measure of the distance from the centre of mass of the M-dwarf to L1, that is a linear measure of its radius relative to the Roche lobe, not a volume-based measurement. Our results are shown in Fig.~\ref{fig:vsini}.

Though a lot of scatter is present in our measurements, in particular around orbital phase 0.5 when the analysed lines from the M-dwarf become much shallower, it can be clearly seen that a low degree of Roche-filling is incompatible with the estimated values. Whereas the reduced $\chi^2$ for the 1.252~$\mu m$ line is similar for models with full Roche-lobe filling and for a radius of 90 per cent of the Roche-lobe radius (4.5 and 6.0, respectively), the value more than doubles for a radius of 80 per cent (13.6), and increases by at least a factor of 5 for 70 and 60 per cent (34 and 62, respectively). Therefore, a radius of less than 80 per cent of the Roche-lobe radius is very unlikely for the M-dwarf.

\section{Discussion}

\subsection{The rates of period change}
\label{sec:nudots}

Our pulse timing analysis builds on previous work \citep{Marsh2016, Stiller2018, Gaibor2020} and further confirms the occurrence of spin-down, now confirmed at the 50-$\sigma$ level through this method. Moreover, we have shown that the spin-down rate has been constant to within the observational uncertainties over a period of 17 years for which there are data. Applying a quadratic term correction based on other measurements to the CRTS times, as well as excluding data taken at phases showing scatter, were crucial for obtaining a good fit to the CRTS data. \citet{Marsh2016} did not take into account the large scatter at some orbital phases, which could explain the discrepancy with more recent measurements.

For the first time, we have also succeeded in obtaining an estimate of the beat frequency derivative via Fourier analysis, obtaining a value in good agreement with results from pulse arrival timing. In order to obtain a good fit to the data, amplitudes had to be fit individually in each run given that the pulse strength is clearly variable with time. In addition, data from our most recent observation for which the strength of consecutive pulses was not consistent with the typical behaviour had to be excluded. Possibly these measures have allowed us to obtain a Fourier fit consistent with the pulse timing analysis when previous attempts have failed.

One puzzling result of our Fourier analysis was a significant difference between our derived beat and spin frequency derivatives when they were left to vary freely, which implies a measurable orbital frequency derivative. Our value of $\dot{\nu}_\mathrm{orbit} = (8.43\pm0.37) \times 10^{-19}$~\hzs is in agreement with the upper limit set by \citet{Stiller2018}, but is more than 20$\sigma$ above the upper limit derived by \citet{Peterson2019}.

It is worth noting that \citet{Peterson2019} remarked that they too obtained a non-null $\dot{\nu}_\mathrm{orbit}$ if the timing uncertainties in their data were not accounted for (which would be unrealistic). As their data included photographic plate measurements, that is indeed a great source of uncertainty, and they point out that it can be up to 11~minutes for their data. Since the inclusion of uncertainties changes what is obtained for $\dot{\nu}_\mathrm{orbit}$, that implies that the significance of $\dot{\nu}_\mathrm{orbit}$ depends on the assumed uncertainties. \citet{Peterson2019}'s conclusion could for example be altered if their uncertainties were overestimated, that is if their timing measurements were more precise than they assumed. The timing uncertainties in our measurements were taken into account, though they are negligible.

\citet{Peterson2019} support their own empirical estimate by pointing out that the models of \citet{Knigge2011} predict $\dot{\nu}_\mathrm{orbit} \lesssim 2 \times 10^{-20}$~\hzs for \arsco's orbital period. However, this prediction does not take into account that \arsco\ is highly magnetic, which affects angular momentum loss \cite[e.g.][]{Kolb1995,Cohen2012,Belloni2020}. The direction in which angular momentum loss is affected, however, is a subject of much discussion. According to \citet{Liebert1985} and \citet{King1985}, the white dwarf magnetic field would lead to enhanced magnetic breaking. They argued that the wind from the donor can be temporarily trapped by the white dwarf's magnetic field lines, thus gaining angular momentum and subsequently carrying more angular momentum out of the system. \citet{Li1994}, on the other hand, proposed the opposite: that the donor's wind would remain trapped in the system due to the white dwarf's magnetic field, leading to reduced angular momentum loss. This model was supported by modelling of the polar population done by \citet{Belloni2020}.

Given that we find a higher $\dot{\nu}_\mathrm{orbit}$ than the upper limit suggested by the non-magnetic semi-empirical model of \citet{Knigge2011}, this would require \arsco\ to be losing more angular momentum than a non-magnetic system, or enhanced magnetic braking. We estimate $\dot{J}_\mathrm{magnetic} \gtrsim 40 \dot{J}_\mathrm{non-magnetic}$ given the ratio between our obtained value and the estimate using \citet{Knigge2011}'s model. Considering the uncertainties surrounding angular momentum loss in magnetic cataclysmic variables, particularly in the case of a unique system such as \arsco, it is hard to establish whether this is feasible or not.

The orbital period change could instead happen at fixed angular momentum if it is triggered by structural changes in the secondary, which can be explained by magnetic activity. Activity can affect the distribution of angular momentum within the star, changing its rotational oblateness (i.e. the gravitational quadrupole moment), which translates into a change in the gravitational field. That in turn causes the orbit and speed of the stars to change, altering the orbital period of the system without affecting its angular momentum \citep[the so-called Applegate mechanism,][]{Applegate1992, Lanza1998, Lanza1999}. This model was proposed to explain the orbital period changes often observed for eclipsing systems \citep[e.g.][]{Parsons2010,Bours2016}, which in many cases are well above the rate expected from magnetic breaking. In addition, the rate of period change caused by the Applegate mechanism is not constant, and can even change in signal (i.e. the period can go from decreasing to increasing and vice-versa). This could potentially explain the discrepancy between our derived value and the limit obtained by \citet{Peterson2019}. This explanation of course requires the M-dwarf to be magnetically active, which seems to be the case at least for \arsco.

%Equation 13 of \citet{Schreiber2021} presents a correction to angular momentum loss due to magnetic breaking in the presence of a strongly magnetised white dwarf:
%\begin{eqnarray}
%    \dot{J}_\mathrm{magnetic} = \dot{J}_\mathrm{non-magnetic} \left( %\frac{\Phi}{0.258} \right)^{5/3}
%\end{eqnarray}
%where $\Phi$ is the fraction of open magnetic field lines.

\subsection{AR~Sco as a cosmic clock}
\label{sec:clock}

\begin{figure}
	\includegraphics[width=\columnwidth]{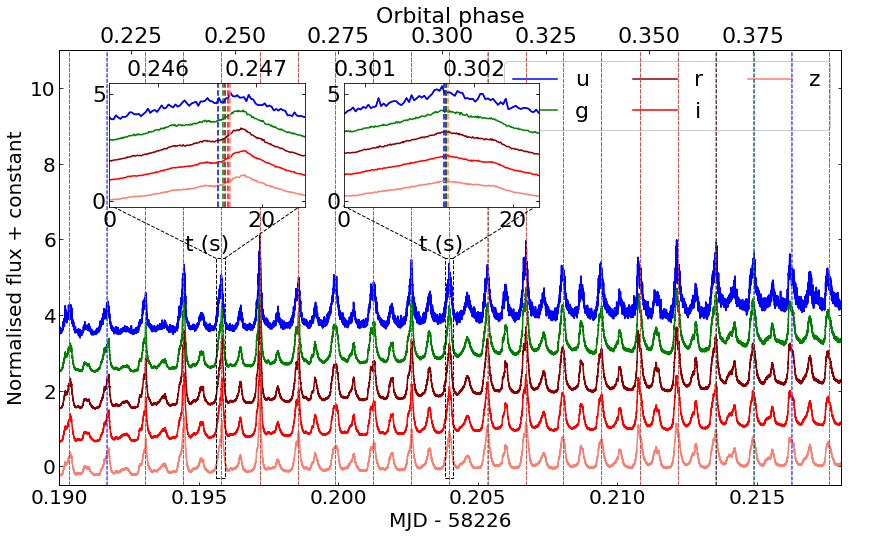}
    \caption{A section of one of our HiPERCAM runs showing data obtained with the $u$, $g$, $r$, $i$, and $z$ filters (from top to bottom). The vertical lines show the pulse arrival times obtained for each filter. In the two insets, it becomes clear that there is a varying level of difference in the pulse arrival times.}
    \label{fig:pulses}
\end{figure}

The regularity and strength of the beat pulsations observed for \arsco, as well as the precise ephemeris that can be obtained for the beat period, raise the question of whether \arsco\ could be used for calibrating the timing accuracy of high-speed observations. This is potentially useful given that it is fairly bright, varies at a speed that it accessible to many instruments, and there is not a need to target particular eclipses. Whereas the ephemerides are precise to less than a second, one important issue is their colour dependency. As already pointed out by \citet{Gaibor2020}, the times of the primary pulses depend on the choice of filter, and their difference varies with orbital phase. This is illustrated in Figure~\ref{fig:pulses} for a section of one of our HiPERCAM runs.

\begin{figure}
	\includegraphics[width=\columnwidth]{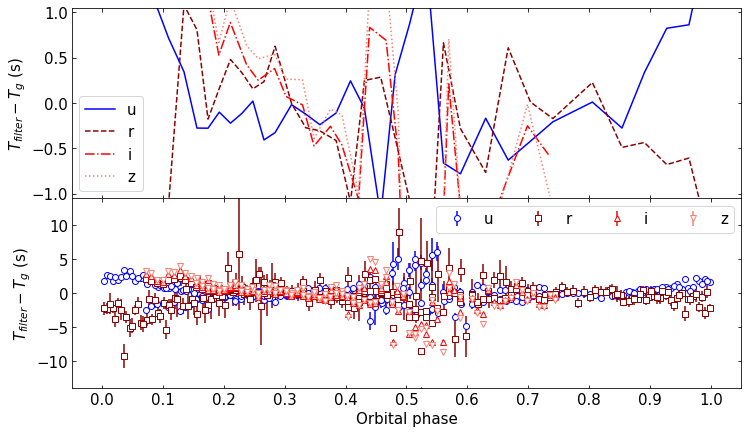}
    \caption{Difference between pulse arrival times for $u,r,i,z$ compared to $g$. The bottom panel shows the individual pulses for our two HiPERCAM runs, as well as one ULTRACAM run added to completely cover the orbital phases. The top panel shows the average every five pulses to illustrate the overall behaviour.}
    \label{fig:chrom}
\end{figure}

It is clear, therefore, that the use of \arsco\ as a timing source depends on the desired accuracy. If the orbital phases around 0.05 and 0.5 are avoided, an accuracy of the order of one second could be obtained fairly easily (see top panel of Figure~\ref{fig:chrom}), although a few pulses times may require combining if signal-to-noise is low. Around phases 0.35 and 0.85, the arrival times for different filters converge, so a better accuracy can be obtained, and one could perhaps hope to calibrate to a few tenths of a second. However, for large telescopes and high signal-to-noise, AR~Sco is not fully reliable as a timing fiducial, and systematic errors from the source itself will limit precision. An overall important result is that, if the timing calibration is being done with a specific filter, preferably an ephemeris obtained with the same filter should be used. Nevertheless, AR~Sco could prove useful in some instances, for example as a quick check against the large errors that can sometimes plague timing work, such as whether the start, middle or end of the exposure has been used, and whether barycentric corrections have been implemented correctly.

\subsection{The behaviour of the companion}

Our Doppler tomography confirms the existence of prominences around the M-dwarf previously identified by \citet{Garnavich2019}. We have further showed that those persist throughout the beat cycle, which suggests that they are not triggered by interaction between the M-dwarf and the white dwarf. Instead they are likely inherent to the M-dwarf, and remain stable due to its magnetic field. Further evidence for this is the fact that similar behaviour is observed for other non-accreting systems \citep[e.g. QS~Vir,][]{Parsons2016}, and even for CVs during low-states \citep[e.g. BB~Dor,][]{Schmidtobreick2012}, and is thus not unique to AR~Sco. Likely the prominences are explained by the magnetic field altering the Roche geometry and displacing the location of the L4 and L5 Lagrange points.

By performing modulated Doppler tomography, we have identified the occurrence of emission modulated on the orbital period, in particular along the line connecting the two stars, but also remarkably in the region where an accretion stream would be located if there were mass transfer. A possible cause for this is a shock wave in the long-lived prominence that is located in this region, triggered by interaction between the two stars' magnetic fields.

In addition, by measuring the $v_\mathrm{rot} \sin i$ for different orbital phases, we provide evidence that the companion is likely filling at least 80 per cent of its Roche-lobe.

%\subsection{UV temperature}
%contribution of white dwarf to the pulses, possible alternative explanation, e.g. hydrogen in the optical path
% role of compressional heating

% compare the spectra light curve (use molly) with the average HST/Kepler light curve

% periodogram

\section{Summary \& Conclusions}

We performed an extensive study of the white dwarf pulsar \arsco\ using precise photometric data spanning almost seven years, complemented by CRTS data that extends our baseline to 2005. We refined the beat period ephemeris using pulse time arrival analysis, detecting a non-zero frequency derivative at the 50$\sigma$ level. We also confirm this value via Fourier analysis. We performed two different fits, first constraining the difference between $\dot{\nu}_\mathrm{spin}$ and $\dot{\nu}_\mathrm{beat}$ to be no more than $3.8 \times 10^{-20}$~\hzs, the $\dot{\nu}_\mathrm{orbit}$ upper limit found by \citet{Peterson2019}, and next leaving both to vary unconstrained. Both cases lead to significant values of $\dot{\nu}_\mathrm{orbit}$, in the first case $(3.68\pm0.17) \times 10^{-20}$~\hzs, and $(8.43\pm0.37) \times 10^{-19}$~\hzs when no prior is applied. This could suggest that \arsco\ is undergoing orbital period changes in excess of what is predicted by magnetic breaking and gravitational wave losses, similar to what is observed for many eclipsing systems and often attributed to changes in the quadrupole moment of the M-dwarf due to magnetic activity. However, since the $\chi^2$ values obtained for our two fits are comparable and they lead to very different $\dot{\nu}_\mathrm{orbit}$ estimates, we believe that there is no strong evidence for measurable orbital period change at the moment, though continuous monitoring might change this picture. There is also no reason to believe $\dot{\nu}_\mathrm{orbit}$ to be constant, as its value can depend on the magnetic activity of the M-dwarf. Large changes have been observed for post-common envelope systems \citep{Parsons2010, Bours2016}.

We additionally carry out an analysis of time-resolved spectra obtained with both ISIS/WHT and X-shooter/VLT. Modulated Doppler tomography, performed here for the first time, shows the occurrence of emission modulated on the orbital period in the region where an accretion stream would be. This region also shows long-lived prominences that appear inherent to the M-dwarf, given their lack of dependence on beat phase. The modulated emission could thus be the result of shock interaction between the two stars' magnetic fields in this region as the system rotates. Our spectra also reveal evidence of tidal deformation detected via the variation on the observed projected rotational velocity, which indicates that the M-dwarf must be at least 80 per cent Roche lobe filling.

\arsco\ remains a unique and puzzling system. If our findings of a possibly significant $\dot{\nu}_\mathrm{orbit}$ are confirmed, they could require the system to be losing angular momentum at a higher rate than non-magnetic systems, which could provide clues to explain the origin of \arsco. Alternatively, this could provide a probe for the mechanism proposed by \citet{Applegate1992}, which allows for orbital period changes at fixed angular momentum. The detection of modulated emission can shed light on the interaction between the M-dwarf and the white dwarf, which remains not fully understood.

\section*{Acknowledgements}

We thank D. Steeghs for advice on how to best visualise modulated Doppler maps. IP and TRM acknowledge support from the UK's Science and Technology Facilities Council (STFC), grant ST/T000406/1. AA acknowledges funding support from the NSRF via the Program Management Unit for Human Resources \& Institutional Development, Research and Innovation, grant B05F640046.
 SGP acknowledges the support of a STFC Ernest Rutherford Fellowship. This work has made use of data obtained at the Thai National Observatory on Doi Inthanon, operated by NARIT, and of observations made with the Gran Telescopio Canarias (GTC), installed at the Spanish Observatorio del Roque de los Muchachos of the Instituto de Astrofísica de Canarias, in the island of La Palma. The design and construction of HiPERCAM was funded by the European Research Council under the European Union’s Seventh Framework Programme (FP/2007-2013) under ERC-2013-ADG Grant Agreement no. 340040 (HiPERCAM). VSD and ULTRACAM/ULTRASPEC/HiPERCAM operations are supported by STFC grant ST/V000853/1. This works is also partially based on observations collected at the European Organisation for Astronomical Research in the Southern Hemisphere under ESO programme 097.D-0685(B), and on observations made with WHT operated on the island of La Palma by the Isaac Newton Group of Telescopes in the Spanish Observatorio del Roque de los Muchachos of the Instituto de Astrof\'{i}sica de Canarias. The ISIS spectroscopy was obtained as part of W/2016A/26.

%%%%%%%%%%%%%%%%%%%%%%%%%%%%%%%%%%%%%%%%%%%%%%%%%%
\section*{Data Availability}

All data analysed in this work can be made available upon reasonable request to the authors.



\bibliographystyle{mnras}
\bibliography{arsco} % if your bibtex file is called example.bib

\begin{thebibliography}{}
\makeatletter
\relax
\def\mn@urlcharsother{\let\do\@makeother \do\$\do\&\do\#\do\^\do\_\do\%\do\~}
\def\mn@doi{\begingroup\mn@urlcharsother \@ifnextchar [ {\mn@doi@}
  {\mn@doi@[]}}
\def\mn@doi@[#1]#2{\def\@tempa{#1}\ifx\@tempa\@empty \href
  {http://dx.doi.org/#2} {doi:#2}\else \href {http://dx.doi.org/#2} {#1}\fi
  \endgroup}
\def\mn@eprint#1#2{\mn@eprint@#1:#2::\@nil}
\def\mn@eprint@arXiv#1{\href {http://arxiv.org/abs/#1} {{\tt arXiv:#1}}}
\def\mn@eprint@dblp#1{\href {http://dblp.uni-trier.de/rec/bibtex/#1.xml}
  {dblp:#1}}
\def\mn@eprint@#1:#2:#3:#4\@nil{\def\@tempa {#1}\def\@tempb {#2}\def\@tempc
  {#3}\ifx \@tempc \@empty \let \@tempc \@tempb \let \@tempb \@tempa \fi \ifx
  \@tempb \@empty \def\@tempb {arXiv}\fi \@ifundefined
  {mn@eprint@\@tempb}{\@tempb:\@tempc}{\expandafter \expandafter \csname
  mn@eprint@\@tempb\endcsname \expandafter{\@tempc}}}

\bibitem[\protect\citeauthoryear{{Applegate}}{{Applegate}}{1992}]{Applegate1992}
{Applegate} J.~H.,  1992, \mn@doi [\apj] {10.1086/170967}, \href
  {https://ui.adsabs.harvard.edu/abs/1992ApJ...385..621A} {385, 621}

\bibitem[\protect\citeauthoryear{{Belloni}, {Schreiber}, {Pala},
  {G{\"a}nsicke}, {Zorotovic}  \& {Rodrigues}}{{Belloni}
  et~al.}{2020}]{Belloni2020}
{Belloni} D.,  {Schreiber} M.~R.,  {Pala} A.~F.,  {G{\"a}nsicke} B.~T.,
  {Zorotovic} M.,   {Rodrigues} C.~V.,  2020, \mn@doi [\mnras]
  {10.1093/mnras/stz3413}, \href
  {https://ui.adsabs.harvard.edu/abs/2020MNRAS.491.5717B} {491, 5717}

\bibitem[\protect\citeauthoryear{{Bours} et~al.,}{{Bours}
  et~al.}{2016}]{Bours2016}
{Bours} M.~C.~P.,  et~al., 2016, \mn@doi [\mnras] {10.1093/mnras/stw1203},
  \href {https://ui.adsabs.harvard.edu/abs/2016MNRAS.460.3873B} {460, 3873}

\bibitem[\protect\citeauthoryear{{Buckley}, {Meintjes}, {Potter}, {Marsh}  \&
  {G{\"a}nsicke}}{{Buckley} et~al.}{2017}]{Buckley2017}
{Buckley} D.~A.~H.,  {Meintjes} P.~J.,  {Potter} S.~B.,  {Marsh} T.~R.,
  {G{\"a}nsicke} B.~T.,  2017, \mn@doi [Nature Astronomy]
  {10.1038/s41550-016-0029}, \href
  {https://ui.adsabs.harvard.edu/abs/2017NatAs...1E..29B} {1, 0029}

\bibitem[\protect\citeauthoryear{{Calvelo}, {Vrtilek}, {Steeghs}, {Torres},
  {Neilsen}, {Filippenko}  \& {Gonz{\'a}lez Hern{\'a}ndez}}{{Calvelo}
  et~al.}{2009}]{Calvelo2009}
{Calvelo} D.~E.,  {Vrtilek} S.~D.,  {Steeghs} D.,  {Torres} M.~A.~P.,
  {Neilsen} J.,  {Filippenko} A.~V.,   {Gonz{\'a}lez Hern{\'a}ndez} J.~I.,
  2009, \mn@doi [\mnras] {10.1111/j.1365-2966.2009.15304.x}, \href
  {https://ui.adsabs.harvard.edu/abs/2009MNRAS.399..539C} {399, 539}

\bibitem[\protect\citeauthoryear{{Cohen}, {Drake}  \& {Kashyap}}{{Cohen}
  et~al.}{2012}]{Cohen2012}
{Cohen} O.,  {Drake} J.~J.,   {Kashyap} V.~L.,  2012, \mn@doi [\apjl]
  {10.1088/2041-8205/746/1/L3}, \href
  {https://ui.adsabs.harvard.edu/abs/2012ApJ...746L...3C} {746, L3}

\bibitem[\protect\citeauthoryear{{Copperwheat}, {Marsh}, {Dhillon},
  {Littlefair}, {Hickman}, {G{\"a}nsicke}  \& {Southworth}}{{Copperwheat}
  et~al.}{2010}]{Copperwheat2010}
{Copperwheat} C.~M.,  {Marsh} T.~R.,  {Dhillon} V.~S.,  {Littlefair} S.~P.,
  {Hickman} R.,  {G{\"a}nsicke} B.~T.,   {Southworth} J.,  2010, \mn@doi
  [\mnras] {10.1111/j.1365-2966.2009.16010.x}, \href
  {https://ui.adsabs.harvard.edu/abs/2010MNRAS.402.1824C} {402, 1824}

\bibitem[\protect\citeauthoryear{{Dhillon} et~al.,}{{Dhillon}
  et~al.}{2007}]{ultracam}
{Dhillon} V.~S.,  et~al., 2007, \mn@doi [\mnras]
  {10.1111/j.1365-2966.2007.11881.x}, \href
  {https://ui.adsabs.harvard.edu/abs/2007MNRAS.378..825D} {378, 825}

\bibitem[\protect\citeauthoryear{{Dhillon} et~al.,}{{Dhillon}
  et~al.}{2014}]{ultraspec}
{Dhillon} V.~S.,  et~al., 2014, \mn@doi [\mnras] {10.1093/mnras/stu1660}, \href
  {https://ui.adsabs.harvard.edu/abs/2014MNRAS.444.4009D} {444, 4009}

\bibitem[\protect\citeauthoryear{{Dhillon} et~al.,}{{Dhillon}
  et~al.}{2021}]{Dhillon2021}
{Dhillon} V.~S.,  et~al., 2021, \mn@doi [\mnras] {10.1093/mnras/stab2130},
  \href {https://ui.adsabs.harvard.edu/abs/2021MNRAS.507..350D} {507, 350}

\bibitem[\protect\citeauthoryear{{Drake} et~al.,}{{Drake}
  et~al.}{2009}]{Drake2009}
{Drake} A.~J.,  et~al., 2009, \mn@doi [\apj] {10.1088/0004-637X/696/1/870},
  \href {https://ui.adsabs.harvard.edu/abs/2009ApJ...696..870D} {696, 870}

\bibitem[\protect\citeauthoryear{{Ferreira}}{{Ferreira}}{2000}]{Ferreira2000}
{Ferreira} J.~M.,  2000, \mn@doi [\mnras] {10.1046/j.1365-8711.2000.03540.x},
  \href {https://ui.adsabs.harvard.edu/abs/2000MNRAS.316..647F} {316, 647}

\bibitem[\protect\citeauthoryear{{Foreman-Mackey}, {Hogg}, {Lang}  \&
  {Goodman}}{{Foreman-Mackey} et~al.}{2013}]{emcee}
{Foreman-Mackey} D.,  {Hogg} D.~W.,  {Lang} D.,   {Goodman} J.,  2013, \mn@doi
  [\pasp] {10.1086/670067}, \href
  {https://ui.adsabs.harvard.edu/abs/2013PASP..125..306F} {125, 306}

\bibitem[\protect\citeauthoryear{{Gaibor}, {Garnavich}, {Littlefield}, {Potter}
   \& {Buckley}}{{Gaibor} et~al.}{2020}]{Gaibor2020}
{Gaibor} Y.,  {Garnavich} P.~M.,  {Littlefield} C.,  {Potter} S.~B.,
  {Buckley} D.~A.~H.,  2020, \mn@doi [\mnras] {10.1093/mnras/staa1901}, \href
  {https://ui.adsabs.harvard.edu/abs/2020MNRAS.496.4849G} {496, 4849}

\bibitem[\protect\citeauthoryear{{Garnavich}, {Littlefield}, {Kafka},
  {Kennedy}, {Callanan}, {Balsara}  \& {Lyutikov}}{{Garnavich}
  et~al.}{2019}]{Garnavich2019}
{Garnavich} P.,  {Littlefield} C.,  {Kafka} S.,  {Kennedy} M.,  {Callanan} P.,
  {Balsara} D.~S.,   {Lyutikov} M.,  2019, \mn@doi [\apj]
  {10.3847/1538-4357/aafb2c}, \href
  {https://ui.adsabs.harvard.edu/abs/2019ApJ...872...67G} {872, 67}

\bibitem[\protect\citeauthoryear{{Garnavich}, {Littlefield}, {Lyutikov}  \&
  {Barkov}}{{Garnavich} et~al.}{2021}]{Garnavich2021}
{Garnavich} P.,  {Littlefield} C.,  {Lyutikov} M.,   {Barkov} M.,  2021,
  \mn@doi [\apj] {10.3847/1538-4357/abd4db}, \href
  {https://ui.adsabs.harvard.edu/abs/2021ApJ...908..195G} {908, 195}

\bibitem[\protect\citeauthoryear{{Geng}, {Zhang}  \& {Huang}}{{Geng}
  et~al.}{2016}]{Geng2016}
{Geng} J.-J.,  {Zhang} B.,   {Huang} Y.-F.,  2016, \mn@doi [\apjl]
  {10.3847/2041-8205/831/1/L10}, \href
  {https://ui.adsabs.harvard.edu/abs/2016ApJ...831L..10G} {831, L10}

\bibitem[\protect\citeauthoryear{{Irawati} et~al.,}{{Irawati}
  et~al.}{2016}]{Irawati2016}
{Irawati} P.,  et~al., 2016, \mn@doi [\mnras] {10.1093/mnras/stv2810}, \href
  {https://ui.adsabs.harvard.edu/abs/2016MNRAS.456.2446I} {456, 2446}

\bibitem[\protect\citeauthoryear{{Isern}, {Garc{\'\i}a-Berro}, {K{\"u}lebi}  \&
  {Lor{\'e}n-Aguilar}}{{Isern} et~al.}{2017}]{Isern2017}
{Isern} J.,  {Garc{\'\i}a-Berro} E.,  {K{\"u}lebi} B.,   {Lor{\'e}n-Aguilar}
  P.,  2017, \mn@doi [\apjl] {10.3847/2041-8213/aa5eae}, \href
  {https://ui.adsabs.harvard.edu/abs/2017ApJ...836L..28I} {836, L28}

\bibitem[\protect\citeauthoryear{{Kafka}, {Ribeiro}, {Baptista}, {Honeycutt}
  \& {Robertson}}{{Kafka} et~al.}{2008}]{Kafka2008}
{Kafka} S.,  {Ribeiro} T.,  {Baptista} R.,  {Honeycutt} R.~K.,   {Robertson}
  J.~W.,  2008, \mn@doi [\apj] {10.1086/592186}, \href
  {https://ui.adsabs.harvard.edu/abs/2008ApJ...688.1302K} {688, 1302}

\bibitem[\protect\citeauthoryear{{Katz}}{{Katz}}{2017}]{Katz2017}
{Katz} J.~I.,  2017, \mn@doi [\apj] {10.3847/1538-4357/835/2/150}, \href
  {https://ui.adsabs.harvard.edu/abs/2017ApJ...835..150K} {835, 150}

\bibitem[\protect\citeauthoryear{{Kausch} et~al.,}{{Kausch}
  et~al.}{2015}]{molecfit2}
{Kausch} W.,  et~al., 2015, \mn@doi [\aap] {10.1051/0004-6361/201423909}, \href
  {https://ui.adsabs.harvard.edu/abs/2015A&A...576A..78K} {576, A78}

\bibitem[\protect\citeauthoryear{{King}}{{King}}{1985}]{King1985}
{King} A.~R.,  1985, \mn@doi [\mnras] {10.1093/mnras/217.1.23P}, \href
  {https://ui.adsabs.harvard.edu/abs/1985MNRAS.217P..23K} {217, 23P}

\bibitem[\protect\citeauthoryear{{Knigge}, {Baraffe}  \& {Patterson}}{{Knigge}
  et~al.}{2011}]{Knigge2011}
{Knigge} C.,  {Baraffe} I.,   {Patterson} J.,  2011, \mn@doi [\apjs]
  {10.1088/0067-0049/194/2/28}, \href
  {https://ui.adsabs.harvard.edu/abs/2011ApJS..194...28K} {194, 28}

\bibitem[\protect\citeauthoryear{{Kolb}}{{Kolb}}{1995}]{Kolb1995}
{Kolb} U.,  1995, in {Buckley} D.~A.~H.,  {Warner} B.,  eds,  Astronomical
  Society of the Pacific Conference Series Vol. 85, Magnetic Cataclysmic
  Variables. p.~440

\bibitem[\protect\citeauthoryear{{Lanza} \& {Rodon{\`o}}}{{Lanza} \&
  {Rodon{\`o}}}{1999}]{Lanza1999}
{Lanza} A.~F.,  {Rodon{\`o}} M.,  1999, \aap, \href
  {https://ui.adsabs.harvard.edu/abs/1999A&A...349..887L} {349, 887}

\bibitem[\protect\citeauthoryear{{Lanza}, {Rodono}  \& {Rosner}}{{Lanza}
  et~al.}{1998}]{Lanza1998}
{Lanza} A.~F.,  {Rodono} M.,   {Rosner} R.,  1998, \mn@doi [\mnras]
  {10.1046/j.1365-8711.1998.01446.x}, \href
  {https://ui.adsabs.harvard.edu/abs/1998MNRAS.296..893L} {296, 893}

\bibitem[\protect\citeauthoryear{{Li}, {Wu}  \& {Wickramasinghe}}{{Li}
  et~al.}{1994}]{Li1994}
{Li} J.~K.,  {Wu} K.~W.,   {Wickramasinghe} D.~T.,  1994, \mn@doi [\mnras]
  {10.1093/mnras/270.4.769}, \href
  {https://ui.adsabs.harvard.edu/abs/1994MNRAS.270..769L} {270, 769}

\bibitem[\protect\citeauthoryear{{Liebert} \& {Stockman}}{{Liebert} \&
  {Stockman}}{1985}]{Liebert1985}
{Liebert} J.,  {Stockman} H.~S.,  1985, in {Lamb} D.~Q.,  {Patterson} J.,  eds,
  Cataclysmic Variables and Low-Mass X-ray Binaries. p.~151,
  \mn@doi{10.1007/978-94-009-5319-2\_20}

\bibitem[\protect\citeauthoryear{{Littlefield}, {Garnavich}, {Kennedy},
  {Callanan}, {Shappee}  \& {Holoien}}{{Littlefield}
  et~al.}{2017}]{Littlefield2017}
{Littlefield} C.,  {Garnavich} P.,  {Kennedy} M.,  {Callanan} P.,  {Shappee}
  B.,   {Holoien} T.,  2017, \mn@doi [\apjl] {10.3847/2041-8213/aa8300}, \href
  {https://ui.adsabs.harvard.edu/abs/2017ApJ...845L...7L} {845, L7}

\bibitem[\protect\citeauthoryear{{Lyutikov}, {Barkov}, {Route}, {Balsara},
  {Garnavich}  \& {Littlefield}}{{Lyutikov} et~al.}{2020}]{Lyutikov2020}
{Lyutikov} M.,  {Barkov} M.,  {Route} M.,  {Balsara} D.,  {Garnavich} P.,
  {Littlefield} C.,  2020, arXiv e-prints, \href
  {https://ui.adsabs.harvard.edu/abs/2020arXiv200411474L} {p. arXiv:2004.11474}

\bibitem[\protect\citeauthoryear{{Marcote}, {Marsh}, {Stanway}, {Paragi}  \&
  {Blanchard}}{{Marcote} et~al.}{2017}]{Marcote2017}
{Marcote} B.,  {Marsh} T.~R.,  {Stanway} E.~R.,  {Paragi} Z.,   {Blanchard}
  J.~M.,  2017, \mn@doi [\aap] {10.1051/0004-6361/201730948}, \href
  {https://ui.adsabs.harvard.edu/abs/2017A&A...601L...7M} {601, L7}

\bibitem[\protect\citeauthoryear{{Marsh}}{{Marsh}}{1989}]{Marsh1989}
{Marsh} T.~R.,  1989, \mn@doi [\pasp] {10.1086/132570}, \href
  {https://ui.adsabs.harvard.edu/abs/1989PASP..101.1032M} {101, 1032}

\bibitem[\protect\citeauthoryear{{Marsh}}{{Marsh}}{2001}]{Marsh2001}
{Marsh} T.~R.,  2001, {Doppler Tomography}.
p.~1

\bibitem[\protect\citeauthoryear{{Marsh}, {Horne}, {Schlegel}, {Honeycutt}  \&
  {Kaitchuck}}{{Marsh} et~al.}{1990}]{Marsh1990}
{Marsh} T.~R.,  {Horne} K.,  {Schlegel} E.~M.,  {Honeycutt} R.~K.,
  {Kaitchuck} R.~H.,  1990, \mn@doi [\apj] {10.1086/169446}, \href
  {https://ui.adsabs.harvard.edu/abs/1990ApJ...364..637M} {364, 637}

\bibitem[\protect\citeauthoryear{{Marsh} et~al.,}{{Marsh}
  et~al.}{2016}]{Marsh2016}
{Marsh} T.~R.,  et~al., 2016, \mn@doi [\nat] {10.1038/nature18620}, \href
  {https://ui.adsabs.harvard.edu/abs/2016Natur.537..374M} {537, 374}

\bibitem[\protect\citeauthoryear{{Papadaki}, {Boffin}  \& {Steeghs}}{{Papadaki}
  et~al.}{2008}]{Papadaki2008}
{Papadaki} C.,  {Boffin} H.~M.~J.,   {Steeghs} D.,  2008, Journal of
  Astronomical Data, \href
  {https://ui.adsabs.harvard.edu/abs/2008JAD....14....2P} {14, 2}

\bibitem[\protect\citeauthoryear{{Parsons} et~al.,}{{Parsons}
  et~al.}{2010}]{Parsons2010}
{Parsons} S.~G.,  et~al., 2010, \mn@doi [\mnras]
  {10.1111/j.1365-2966.2010.17063.x}, \href
  {https://ui.adsabs.harvard.edu/abs/2010MNRAS.407.2362P} {407, 2362}

\bibitem[\protect\citeauthoryear{{Parsons} et~al.,}{{Parsons}
  et~al.}{2016}]{Parsons2016}
{Parsons} S.~G.,  et~al., 2016, \mn@doi [\mnras] {10.1093/mnras/stw516}, \href
  {https://ui.adsabs.harvard.edu/abs/2016MNRAS.458.2793P} {458, 2793}

\bibitem[\protect\citeauthoryear{{Parsons} et~al.,}{{Parsons}
  et~al.}{2018}]{Parsons2018}
{Parsons} S.~G.,  et~al., 2018, \mn@doi [\mnras] {10.1093/mnras/sty2345}, \href
  {https://ui.adsabs.harvard.edu/abs/2018MNRAS.481.1083P} {481, 1083}

\bibitem[\protect\citeauthoryear{{Patterson} et~al.,}{{Patterson}
  et~al.}{2020}]{Patterson2020}
{Patterson} J.,  et~al., 2020, \mn@doi [\apj] {10.3847/1538-4357/ab863d}, \href
  {https://ui.adsabs.harvard.edu/abs/2020ApJ...897...70P} {897, 70}

\bibitem[\protect\citeauthoryear{{Peterson}, {Littlefield}  \&
  {Garnavich}}{{Peterson} et~al.}{2019}]{Peterson2019}
{Peterson} E.,  {Littlefield} C.,   {Garnavich} P.,  2019, \mn@doi [\aj]
  {10.3847/1538-3881/ab2ad5}, \href
  {https://ui.adsabs.harvard.edu/abs/2019AJ....158..131P} {158, 131}

\bibitem[\protect\citeauthoryear{{Potter} \& {Buckley}}{{Potter} \&
  {Buckley}}{2018a}]{PotterBuckley2018Fourier}
{Potter} S.~B.,  {Buckley} D. A.~H.,  2018a, \mn@doi [\mnras]
  {10.1093/mnrasl/sly078}, \href
  {https://ui.adsabs.harvard.edu/abs/2018MNRAS.478L..78P} {478, L78}

\bibitem[\protect\citeauthoryear{{Potter} \& {Buckley}}{{Potter} \&
  {Buckley}}{2018b}]{PotterBuckley2018}
{Potter} S.~B.,  {Buckley} D. A.~H.,  2018b, \mn@doi [\mnras]
  {10.1093/mnras/sty2407}, \href
  {https://ui.adsabs.harvard.edu/abs/2018MNRAS.481.2384P} {481, 2384}

\bibitem[\protect\citeauthoryear{{Schmidtobreick}, {Rodr{\'\i}guez-Gil},
  {Long}, {G{\"a}nsicke}, {Tappert}  \& {Torres}}{{Schmidtobreick}
  et~al.}{2012}]{Schmidtobreick2012}
{Schmidtobreick} L.,  {Rodr{\'\i}guez-Gil} P.,  {Long} K.~S.,  {G{\"a}nsicke}
  B.~T.,  {Tappert} C.,   {Torres} M.~A.~P.,  2012, \mn@doi [\mnras]
  {10.1111/j.1365-2966.2012.20653.x}, \href
  {https://ui.adsabs.harvard.edu/abs/2012MNRAS.422..731S} {422, 731}

\bibitem[\protect\citeauthoryear{{Schreiber}, {Belloni}, {G{\"a}nsicke},
  {Parsons}  \& {Zorotovic}}{{Schreiber} et~al.}{2021}]{Schreiber2021}
{Schreiber} M.~R.,  {Belloni} D.,  {G{\"a}nsicke} B.~T.,  {Parsons} S.~G.,
  {Zorotovic} M.,  2021, \mn@doi [Nature Astronomy]
  {10.1038/s41550-021-01346-8}, \href
  {https://ui.adsabs.harvard.edu/abs/2021NatAs...5..648S} {5, 648}

\bibitem[\protect\citeauthoryear{{Smette} et~al.,}{{Smette}
  et~al.}{2015}]{molecfit1}
{Smette} A.,  et~al., 2015, \mn@doi [\aap] {10.1051/0004-6361/201423932}, \href
  {https://ui.adsabs.harvard.edu/abs/2015A&A...576A..77S} {576, A77}

\bibitem[\protect\citeauthoryear{{Somero}, {Hakala}, {Muhli}, {Charles}  \&
  {Vilhu}}{{Somero} et~al.}{2012}]{Somero2012}
{Somero} A.,  {Hakala} P.,  {Muhli} P.,  {Charles} P.,   {Vilhu} O.,  2012,
  \mn@doi [\aap] {10.1051/0004-6361/201118439}, \href
  {https://ui.adsabs.harvard.edu/abs/2012A&A...539A.111S} {539, A111}

\bibitem[\protect\citeauthoryear{{Stanway}, {Marsh}, {Chote}, {G{\"a}nsicke},
  {Steeghs}  \& {Wheatley}}{{Stanway} et~al.}{2018}]{Stanway2018}
{Stanway} E.~R.,  {Marsh} T.~R.,  {Chote} P.,  {G{\"a}nsicke} B.~T.,  {Steeghs}
  D.,   {Wheatley} P.~J.,  2018, \mn@doi [\aap] {10.1051/0004-6361/201732380},
  \href {https://ui.adsabs.harvard.edu/abs/2018A&A...611A..66S} {611, A66}

\bibitem[\protect\citeauthoryear{{Steeghs}}{{Steeghs}}{2003}]{Steeghs2003}
{Steeghs} D.,  2003, \mn@doi [\mnras] {10.1046/j.1365-8711.2003.06917.x}, \href
  {https://ui.adsabs.harvard.edu/abs/2003MNRAS.344..448S} {344, 448}

\bibitem[\protect\citeauthoryear{{Stiller}, {Littlefield}, {Garnavich}, {Wood},
  {Hambsch}  \& {Myers}}{{Stiller} et~al.}{2018}]{Stiller2018}
{Stiller} R.~A.,  {Littlefield} C.,  {Garnavich} P.,  {Wood} C.,  {Hambsch}
  F.-J.,   {Myers} G.,  2018, \mn@doi [\aj] {10.3847/1538-3881/aad5dd}, \href
  {https://ui.adsabs.harvard.edu/abs/2018AJ....156..150S} {156, 150}

\bibitem[\protect\citeauthoryear{{Takata}, {Yang}  \& {Cheng}}{{Takata}
  et~al.}{2017}]{Takata2017}
{Takata} J.,  {Yang} H.,   {Cheng} K.~S.,  2017, \mn@doi [\apj]
  {10.3847/1538-4357/aa9b33}, \href
  {https://ui.adsabs.harvard.edu/abs/2017ApJ...851..143T} {851, 143}

\bibitem[\protect\citeauthoryear{{Takata}, {Hu}, {Lin}, {Tam}, {Pal}, {Hui},
  {Kong}  \& {Cheng}}{{Takata} et~al.}{2018}]{Takata2018}
{Takata} J.,  {Hu} C.~P.,  {Lin} L.~C.~C.,  {Tam} P.~H.~T.,  {Pal} P.~S.,
  {Hui} C.~Y.,  {Kong} A.~K.~H.,   {Cheng} K.~S.,  2018, \mn@doi [\apj]
  {10.3847/1538-4357/aaa23d}, \href
  {https://ui.adsabs.harvard.edu/abs/2018ApJ...853..106T} {853, 106}

\bibitem[\protect\citeauthoryear{{Vernet} et~al.,}{{Vernet}
  et~al.}{2011}]{Vernet2011}
{Vernet} J.,  et~al., 2011, \mn@doi [\aap] {10.1051/0004-6361/201117752}, \href
  {https://ui.adsabs.harvard.edu/abs/2011A&A...536A.105V} {536, A105}

\bibitem[\protect\citeauthoryear{{du Plessis}, {Wadiasingh}, {Venter}  \&
  {Harding}}{{du Plessis} et~al.}{2019}]{Plessis2019}
{du Plessis} L.,  {Wadiasingh} Z.,  {Venter} C.,   {Harding} A.~K.,  2019,
  \mn@doi [\apj] {10.3847/1538-4357/ab4e19}, \href
  {https://ui.adsabs.harvard.edu/abs/2019ApJ...887...44D} {887, 44}

\bibitem[\protect\citeauthoryear{{du Plessis}, {Venter}, {Wadiasingh},
  {Harding}, {Buckley}, {Potter}  \& {Meintjes}}{{du Plessis}
  et~al.}{2022}]{Plessis2022}
{du Plessis} L.,  {Venter} C.,  {Wadiasingh} Z.,  {Harding} A.~K.,  {Buckley}
  D. A.~H.,  {Potter} S.~B.,   {Meintjes} P.~J.,  2022, \mn@doi [\mnras]
  {10.1093/mnras/stab3595}, \href
  {https://ui.adsabs.harvard.edu/abs/2022MNRAS.510.2998D} {510, 2998}

\makeatother
\end{thebibliography}

%\begin{thebibliography}{99}
%\end{thebibliography}

%%%%%%%%%%%%%%%%%%%%%%%%%%%%%%%%%%%%%%%%%%%%%%%%%%

%%%%%%%%%%%%%%%%% APPENDICES %%%%%%%%%%%%%%%%%%%%%

\appendix

\section{Log of photomometric observations}

\begin{table*}
	\centering
	\caption{Journal of observations.}
    	\label{tab:observations}
        \begin{tabular}{ccccc}
        \hline
        Telescope & Start & Duration & Cadence & Filter \\
                  & (TDB) & (min)    &     (s) &        \\
        \hline
        ULTRACAM & 2015-06-23 21:12:56.852 & 37.7 & 2.9 & $u,g,r$ \\
        ULTRACAM & 2015-06-24 21:12:19.723 & 165.6 & 1.3 & $u,g,i$ \\
        ULTRASPEC & 2016-01-19 22:03:32.622 & 70.2 & 4.0 & $g$ \\
        ULTRASPEC & 2016-03-07 22:39:59.884 & 23.8 & 4.0 & $g$ \\
        ULTRASPEC & 2016-03-08 22:19:40.231 & 42.6 & 4.0 & $g$ \\
        ULTRASPEC & 2016-03-13 21:32:59.534 & 87.4 & 4.7 & $g$ \\
        ULTRASPEC & 2016-03-16 21:24:04.987 & 91.6 & 4.7 & $g$ \\
        ULTRASPEC & 2016-03-17 20:19:08.410 & 45.8 & 4.7 & $g$ \\
        ULTRASPEC & 2016-03-19 19:11:59.534 & 26.6 & 4.7 & $g$ \\
        ULTRASPEC & 2016-03-21 22:09:02.976 & 48.4 & 4.0 & $g$ \\
        ULTRASPEC & 2016-04-08 21:45:16.514 & 60.2 & 4.0 & $g$ \\
        ULTRASPEC & 2016-04-09 21:44:43.571 & 53.3 & 4.0 & $g$ \\
        ULTRASPEC & 2016-04-10 21:54:49.403 & 46.5 & 4.0 & $g$ \\
        ULTRACAM & 2016-04-21 06:23:46.060 & 104.7 & 2.0 & $u,g,r$ \\
        ULTRACAM & 2016-04-21 08:40:39.584 & 117.5 & 2.0 & $u,g,r$ \\
        ULTRASPEC & 2016-05-02 19:44:55.657 & 158.2 & 4.0 & $g$ \\
        ULTRASPEC & 2016-05-07 18:39:43.968 & 3.7 & 3.4 & $g$ \\
        ULTRASPEC & 2016-05-07 18:43:36.310 & 14.6 & 3.4 & $g$ \\
        ULTRASPEC & 2016-05-07 18:58:39.648 & 201.6 & 3.4 & $g$ \\
        ULTRASPEC & 2016-05-08 17:11:27.304 & 311.0 & 5.4 & $g$ \\
        ULTRACAM & 2016-07-04 01:15:02.419 & 13.0 & 2.3 & $u,g,r$ \\
        ULTRACAM & 2016-07-04 01:28:09.289 & 14.9 & 3.0 & $u,g,r$ \\
        ULTRACAM & 2016-07-04 01:43:10.239 & 9.8 & 2.0 & $u,g,r$ \\
        ULTRACAM & 2016-07-04 01:53:06.634 & 59.6 & 2.0 & $u,g,r$ \\
        ULTRACAM & 2016-07-04 02:52:45.285 & 195.3 & 1.0 & $u,g,r$ \\
        ULTRACAM & 2016-07-04 06:08:10.125 & 48.7 & 1.6 & $u,g,r$ \\
        ULTRACAM & 2016-08-19 23:53:04.573 & 62.5 & 4.0 & $u,g,i$ \\
        ULTRASPEC & 2017-01-23 22:13:02.715 & 53.7 & 4.9 & $g$ \\
        ULTRASPEC & 2017-02-10 21:43:24.910 & 39.5 & 6.2 & $g$ \\
        ULTRASPEC & 2017-02-11 22:15:57.551 & 53.0 & 4.2 & $g$ \\
        ULTRASPEC & 2017-02-12 21:21:12.738 & 111.3 & 4.4 & $g$ \\
        ULTRASPEC & 2017-02-14 22:05:13.713 & 65.1 & 4.2 & $g$ \\
        ULTRASPEC & 2017-02-18 21:30:20.061 & 68.5 & 4.5 & $g$ \\
        ULTRASPEC & 2017-02-24 19:59:18.793 & 55.2 & 4.5 & $g$ \\
        ULTRACAM & 2017-03-17 06:39:39.180 & 74.1 & 3.0 & $u,g,r$ \\
        ULTRACAM & 2017-03-21 08:37:59.580 & 61.1 & 3.0 & $u,g,r$ \\
        ULTRASPEC & 2017-03-30 20:47:27.827 & 116.2 & 4.2 & $g$ \\
        ULTRASPEC & 2017-03-31 20:44:02.680 & 111.9 & 4.2 & $g$ \\
        ULTRASPEC & 2017-04-01 22:10:08.743 & 11.2 & 4.2 & $g$ \\
        ULTRASPEC & 2017-04-06 18:07:22.546 & 74.2 & 4.2 & $g$ \\
        ULTRASPEC & 2017-04-07 22:14:25.609 & 25.9 & 4.2 & $g$ \\
        ULTRASPEC & 2017-04-08 21:20:50.174 & 77.0 & 4.2 & $g$ \\
        ULTRACAM & 2017-05-03 03:00:53.854 & 1.3 & 3.0 & $u,g,i$ \\
        ULTRACAM & 2017-05-03 03:02:22.843 & 127.2 & 3.0 & $u,g,i$ \\
        ULTRACAM & 2017-05-04 01:43:27.069 & 219.5 & 3.0 & $u,g,r$ \\
        ULTRACAM & 2017-05-04 05:23:03.800 & 312.2 & 1.3 & $u,g,r$ \\
        ULTRACAM & 2017-05-05 01:48:11.118 & 529.9 & 1.3 & $u,g,r$ \\
        ULTRACAM & 2017-05-06 06:01:02.773 & 174.2 & 1.3 & $u,g,r$ \\
        ULTRACAM & 2017-05-14 06:15:24.989 & 268.6 & 3.0 & $u_s,g_s,r_s$ \\
        ULTRACAM & 2017-06-10 02:26:04.269 & 223.9 & 1.3 & $u_s,g_s,r_s$ \\
        ULTRACAM & 2018-01-19 07:58:18.422 & 77.4 & 2.0 & $u_s,g_s,r_s$ \\
        ULTRACAM & 2018-01-20 08:05:59.798 & 60.2 & 1.3 & $u_s,g_s,r_s$ \\
        ULTRACAM & 2018-01-22 07:57:58.542 & 72.3 & 1.3 & $u_s,g_s,r_s$ \\
        ULTRASPEC & 2018-03-03 20:06:08.847 & 154.4 & 3.4 & $g$ \\
        ULTRASPEC & 2018-03-03 22:40:41.706 & 17.1 & 3.4 & $g$ \\
        ULTRASPEC & 2018-03-04 19:57:44.470 & 160.1 & 4.2 & $g$ \\
        ULTRASPEC & 2018-03-04 22:38:00.830 & 14.3 & 4.2 & $g$ \\
        ULTRASPEC & 2018-03-05 19:54:03.936 & 184.4 & 4.2 & $g$ \\
        ULTRASPEC & 2018-03-18 22:15:48.145 & 34.2 & 4.2 & $g$ \\
        ULTRASPEC & 2018-03-19 21:01:17.871 & 36.8 & 4.2 & $g$ \\
        ULTRASPEC & 2018-03-19 21:38:14.630 & 76.6 & 4.2 & $g$ \\
        ULTRASPEC & 2018-03-20 21:54:01.845 & 50.4 & 4.2 & $g$ \\
        \hline
    \end{tabular}
\end{table*}
\begin{table*}
	\centering
	\contcaption{}
    	%\label{tab:observations}
        \begin{tabular}{ccccc}
        \hline
        Telescope & Start & Duration & Cadence & Filter \\
                  & (TDB) & (min)    &     (s) &        \\
        \hline
        ULTRACAM & 2018-04-14 09:44:25.367 & 46.8 & 2.0 & $u_s,g_s,r_s$ \\
        ULTRACAM & 2018-04-15 08:07:05.188 & 45.4 & 1.3 & $u_s,g_s,r_s$ \\
        HiPERCAM & 2018-04-16 02:25:42.704 & 131.0 & 0.1 & $u_s,g_s,r_s,i_s,z_s$ \\
        HiPERCAM & 2018-04-18 04:03:15.289 & 114.8 & 0.1 & $u_s,g_s,r_s,i_s,z_s$ \\
        ULTRACAM & 2018-06-06 05:15:00.776 & 22.7 & 2.0 & $u_s,g_s,r_s$ \\
        ULTRACAM & 2018-06-06 05:38:21.907 & 62.3 & 2.0 & $u_s,g_s,r_s$ \\
        ULTRACAM & 2018-06-19 01:12:53.833 & 177.2 & 2.0 & $u_s,g_s,r_s$ \\
        ULTRASPEC & 2019-02-07 21:08:04.697 & 111.7 & 4.5 & $g$ \\
        ULTRACAM & 2019-03-23 06:53:51.419 & 207.1 & 1.3 & $u_s,g_s,r_s$ \\
        ULTRASPEC & 2020-02-24 20:22:41.377 & 22.5 & 9.1 & $g$ \\
        ULTRASPEC & 2020-03-13 19:31:13.103 & 90.2 & 4.5 & $g$ \\
        ULTRASPEC & 2020-03-23 20:56:00.406 & 27.1 & 4.5 & KG5 \\
        ULTRASPEC & 2020-03-23 21:27:07.128 & 52.8 & 4.5 & KG5 \\
        ULTRASPEC & 2020-03-28 19:56:59.783 & 133.3 & 4.2 & $g$ \\
        ULTRASPEC & 2020-04-03 18:43:20.081 & 234.0 & 4.2 & $r$ \\
        ULTRASPEC & 2020-04-04 19:56:42.781 & 156.0 & 4.2 & $g$ \\
        ULTRASPEC & 2021-02-28 21:00:48.318 & 46.0 & 4.5 & $g$ \\
        ULTRASPEC & 2021-03-01 20:38:24.340 & 74.1 & 4.4 & $g$ \\
        ULTRACAM & 2021-04-08 08:35:44.966 & 40.2 & 2.0 & $u_s,g_s,r_s$ \\
        ULTRASPEC & 2022-02-20 21:17:17.315 & 38.1 & 4.5 & $g$ \\
        ULTRACAM & 2022-04-26 03:25:28.705 & 229.6 & 2.5 & $u_s,g_s,i_s$ \\
        \hline
    \end{tabular}
\end{table*}

\section{Corner plot for the MCMC fit}

\begin{figure*}
	\includegraphics[width=\textwidth]{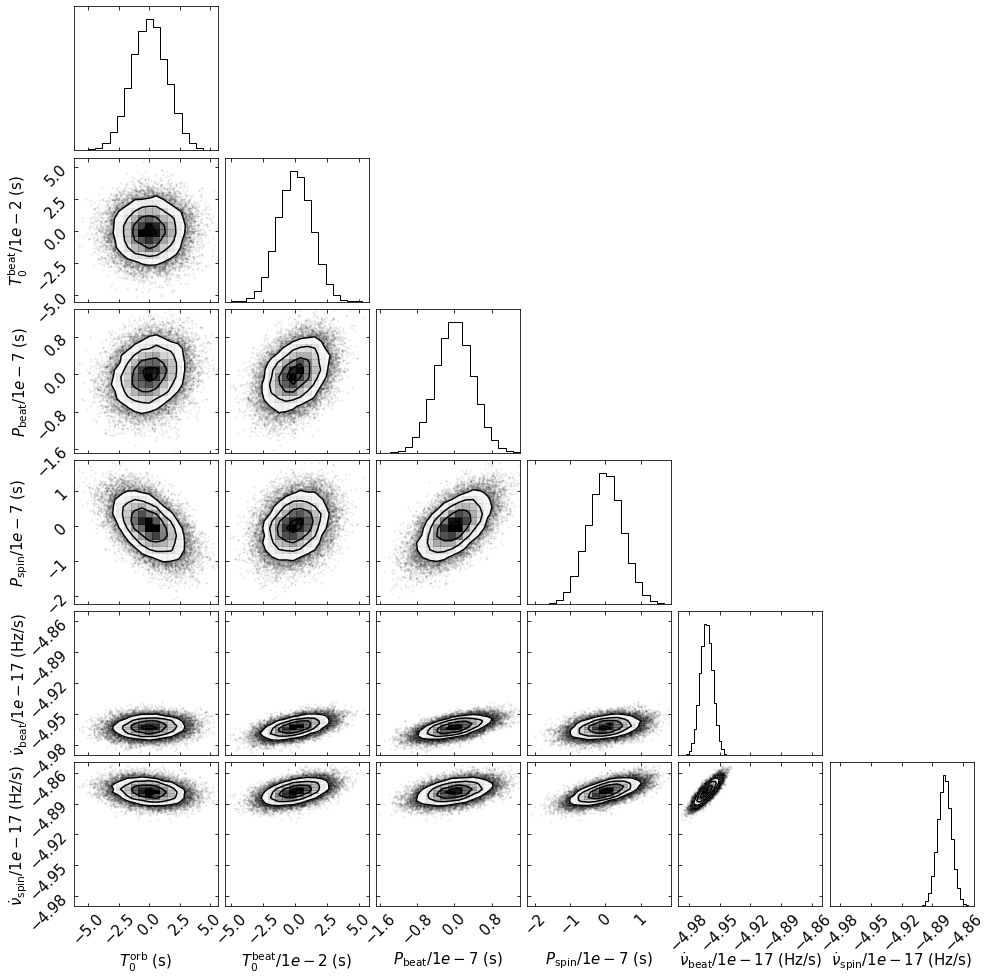}
    \caption{Corner plot resulting from the unconstrained MCMC fit to the data. The median values quoted in the text have been subtracted for both $T_0$ and period values. We show $\dot{\nu}_\mathrm{beat}$ and $\dot{\nu}_\mathrm{beat}$ with the same scale in both axes to facilitate comparison}.
    \label{fig:corner}
\end{figure*}


%%%%%%%%%%%%%%%%%%%%%%%%%%%%%%%%%%%%%%%%%%%%%%%%%%

%\begin{figure}
%	\includegraphics[width=\columnwidth]{example}
%    \caption{This is an example figure. Captions appear below each figure.
%	Give enough detail for the reader to understand what they're looking at,
%	but leave detailed discussion to the main body of the text.}
%    \label{fig:example_figure}
%\end{figure}

%\begin{table}
%	\centering
%	\caption{This is an example table. Captions appear above each table.
%	Remember to define the quantities, symbols and units used.}
%	\label{tab:example_table}
%	\begin{tabular}{lccr} % four columns, alignment for each
%		\hline
%		A & B & C & D\\
%		\hline
%		1 & 2 & 3 & 4\\
%		2 & 4 & 6 & 8\\
%		3 & 5 & 7 & 9\\
%		\hline
%	\end{tabular}
%\end{table}

% Don't change these lines
\bsp	% typesetting comment
\label{lastpage}
\end{document}